\documentclass{aa}  
\usepackage{graphicx}
\usepackage[colorlinks=true,citecolor=blue]{hyperref}
\usepackage{txfonts}
\usepackage{newtxtext, newtxmath}
\usepackage[T1]{fontenc}
\usepackage{amsmath}
\usepackage{amssymb}
\usepackage{color}
\usepackage{xcolor}
\usepackage{hyperref}
\usepackage{lscape}
\usepackage{multirow}
\usepackage{threeparttable}
\usepackage{longtable}
\usepackage{makecell}
\usepackage{bm}
\usepackage{arydshln}
\usepackage{float} 
\usepackage{booktabs} 
\usepackage{diagbox}
\usepackage{array}
\usepackage{lscape}
\usepackage{comment}
\usepackage{multibib}
\newcites{supp}{Additional References}

\newcolumntype{C}[1]{>{\centering\let\newline\\\arraybackslash\hspace{0pt}}m{#1}}

\newcommand{\kms}{\,km\,s$^{-1}$} 
\newcommand{\kmsd}{\,km\,s$^{-1}$\,d$^{-1}$} 
\newcommand{\naid}{\ion{Na}{i}\,D\,}

\newcommand{\caii}{\ion{Ca}{ii}\,}
\newcommand{\ki}{\ion{K}{i}\,}
\newcommand{\sii}{\ion{Si}{ii}\,}

\newcommand{\subCh}{sub-$M_{\mathrm{Ch}}$}
\newcommand{\Ch}{$M_{\mathrm{Ch}}$}

\usepackage{hyperref}
\definecolor{yaleblue}{rgb}{0.1,0.3,0.9}
\definecolor{lava}{rgb}{0.81, 0.06, 0.13}
\definecolor{forestgreen}{rgb}{0.0, 0.45, 0.13}
\hypersetup{colorlinks=true, linkcolor=lava, urlcolor=forestgreen, citecolor=yaleblue}

\newcommand{\orcid}[1]{\href{https://orcid.org/#1}{\includegraphics[width=10pt]{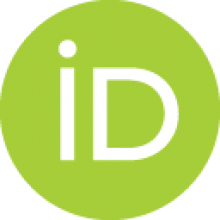}}}

\defcitealias{GG24}{Paper I}
\defcitealias{GG25}{Paper II}
\defcitealias{G25}{Paper III}
\defcitealias{Forster13}{F13}
\begin{document} 

\title{Narrow absorption lines from intervening material in supernovae}

\subtitle{IV. Type Ia supernovae: \naid\ line strength relating to external material and intrinsic properties}

\author{
Santiago Gonz\'alez-Gait\'an
\inst{1}\orcid{0000-0001-9541-0317}
\and
Claudia P. Guti\'errez
\inst{2,3}\orcid{0000-0003-2375-2064}
\and
João Duarte \inst{4}
\and
Rita Santos \inst{4}
\and
Gonçalo Martins \inst{4}
\and
Joseph P. Anderson
\inst{5}\orcid{0000-0003-0227-3451}
\and
Llu\'is Galbany
\inst{3,2}\orcid{0000-0002-1296-6887}
}

\institute{
Instituto de Astrof\'isica e Ci\^encias do Espaço, Faculdade de Ci\^encias, Universidade de Lisboa, Ed. C8, Campo Grande, 1749-016 Lisbon, Portugal\\
\email{gongsale@gmail.com}
\and
Institut d'Estudis Espacials de Catalunya (IEEC), Edifici RDIT, Campus UPC, 08860 Castelldefels (Barcelona), Spain\\
\email{cgutierrez@ice.csic.es}
\and
Institute of Space Sciences (ICE, CSIC), Campus UAB, Carrer de Can Magrans, s/n, E-08193 Barcelona, Spain
\and
CENTRA, Instituto Superior Técnico, Universidade de Lisboa, Av. Rovisco Pais 1, 1049-001 Lisboa, Portugal
\and
European Southern Observatory, Alonso de C\'ordova 3107, Casilla 19, Santiago, Chile
}
\date{}

\abstract
{
Type Ia supernovae (SNe~Ia) are thermonuclear runaways of some white dwarfs in binary systems. They have been extensively studied, yet their progenitor and explosion mechanisms remain poorly understood. We study a large sample of SNe~Ia comparing the narrow interstellar absorption features in their spectra with various photometric and spectroscopic supernova properties, as well as environmental characteristics. The sodium absorption is significantly stronger in younger, more star-forming and more centrally located SNe~Ia, as expected. However, we also show that there is a relation with intrinsic properties that is independent of the environment. In fact, there is substantial evidence for two environmental SN~Ia populations, an old and a young one, with the young population showing significantly different distributions of sodium strength when divided according to the \sii\ ejecta velocity, nebular velocity, extinction, $E(B-V)$, and reddening curve, $R_V$. Performing a clustering of the SNe~Ia, we recover an old population of SNe with low extinction and normal ejecta velocity, while the young population can indeed be subdivided into a group of highly-extincted, high-velocity SNe~Ia with much stronger blueshifted sodium absorption, and another of low-extincted, normal-velocity objects with little sodium absorption. We interpret this relation of intervening material with intrinsic properties as evidence for the young SN~Ia population, occurring in young and star-forming environments, to have asymmetric radiation that interacts with nearby material, and whose observables depend on the viewing angle. Finally, we show that the cosmological mass-step is consistent with these populations.}

\authorrunning{Gonz\'alez-Gait\'an, Guti\'errez et al.}
\titlerunning{SN narrow lines in SNe~Ia}
\maketitle
%

\section{Introduction}

Type Ia supernovae (SNe~Ia) are fascinating bright explosions that mark the end stages of some binary stellar systems, enrich the Universe with heavy elements \citep[e.g.,][]{Matteucci01,Cavichia24}, and are routinely used to measure distances in cosmology \citep[e.g.,][]{Riess98,Perlmutter99,DES24}. Although there is consensus that they originate from the thermonuclear runway of a carbon-oxygen white dwarf \citep[WD; e.g.,][]{Nugent11,Bloom12}, many questions remain on the nature of the companion star and the explosion process (see \citealt{Ruiter25} for a review). 

Progenitor systems have traditionally been subdivided into single-degenerate (SD) systems, in which the companion is a non-degenerate main-sequence, helium or red giant star \citep{Whelan73}, and double-degenerate (DD) scenarios, in which the companion is another degenerate WD \citep{Iben84}; but alternative progenitors have been proposed \citep[e.g.,][]{Wang17,Soker23}. The main explosion scenarios of SNe~Ia are: i) the Chandrasekhar-mass (\Ch) model in which the WD nears the Chandraskehar limit of 1.4$M_{\odot}$ leading to a deflagration at the centre that transitions into a detonation \citep[delayed-detonation,][]{Nomoto84,Khokhlov91}, and ii) the sub-Chandrasekhar (\subCh) model in which the WD accretes helium from a companion that detonates in the outer shell leading to a sub-$M_{\mathrm{Ch}}$ detonation in the core \citep[double-detonation or DDet,][]{Taam80,Nomoto80}. More recent studies have shown that a smaller helium shell at ignition agrees better with observations \citep{Bildsten07,Shen09}, motivating numerous further studies \citep[e.g.,][]{Kromer10,Townsley12,Shen21}. Theoretical and observational considerations have shown that multiple mechanisms may combine to produce the overall "normal" SN~Ia population \citep[e.g.,][]{Ruiter09,Scalzo14,Dhawan17,Cikota26}, without even considering outlying peculiar objects \citep[e.g.,][]{Filippenko92a,Phillips92,Li03}. 

The main mechanism powering the light-curve of SNe~Ia is the radioactive decay of nickel, and successful models need to be able to explain, among other observations, the relation between maximum brightness and light-curve width \citep{Phillips93}, which is also one of the corrections for distance calibration. In addition to the presence of both iron-group elements and intermediate-mass elements in the spectra, models must be able to reproduce a range of ejecta velocities and absorption strengths \citep{Branch06,Wang09}. 

Another part of the puzzle of SN~Ia progenitors is the possible presence of detached material near the explosion. Some SNe~Ia show clear evidence for this material through narrow hydrogen emission lines in their spectra \citep{Hamuy03,Dilday12,Sharma23}, infrared excess and radio emission \citep{Mo25,Kool23}, light-echoes \citep[e.g.,][]{Wang08,Drozdov16}, varying narrow absorption features \citep[e.g.,][]{Patat07,Blondin09,Sternberg14,Ferretti16} that are blueshifted \citep[e.g.,][]{Sternberg11,Maguire13,Phillips13}, among others. If such matter is circumstellar material (CSM) produced prior to the explosion, either from accretion or merger events, this provides key information on the progenitor system \citep[e.g.,][]{Piro16,Moriya23,Inoue26}. If it comes from close-by interstellar clouds, then this has important implications for the physics of the interstellar medium (ISM). 

Associated with the question of nearby material, dust extinction in the line of sight (LoS) of SNe~Ia has been a continuously active subject of debate. As the amount of dust changes from object to object, the reddening alters the observed colours and the associated extinction that dims the light. This explains the linear trend observed in the brightness-colour relation of SNe~Ia, although it is expected both theoretically and observationally that intrinsic colours vary \citep[e.g.,][]{Burns14,Hoeflich17}. To complicate things further, a few SNe~Ia present very peculiar reddening curves \citep[e.g.][]{Foley14,Gutierrez16} and possible variations with environments \citep{Ramaiya25,Duarte25}. Moreover, there is evidence that a non-negligible fraction of SNe~Ia suffer variable extinction \citep{Forster13,Bulla18}. These observations could pose problems to the simple linear brightness-colour relation, i.e. the second correction of the distance calibration  \citep{Tripp98}. In fact, a third brightness correction related to the environment of the SNe~Ia \citep["mass-step", see e.g.][]{Sullivan10,Kelly10} seems to be related to dust and/or intrinsic colours \citep[e.g.,][]{Brout21,GG21}.

In this paper, we analyse a large sample of SNe~Ia focusing on the narrow absorption spectral lines of intervening material in the LoS and compare their equivalent width and velocity to environmental and supernova properties, to shed more light on the ongoing question of the progenitors, the nearby material and the colours and extinction of SNe~Ia. This is the fourth paper of a series: in the first we present the dataset comprising all SN types and the automated techniques to measure the lines, as well as a study of their evolution with time \citep[][hereafter Paper I]{GG24}, in the second we compare the narrow lines to environmental properties \citep[][hereafter Paper II]{GG25}, and in the third we analyse differences among SN types \citep[][hereafter Paper III]{G25}. This paper, dedicated to SNe~Ia, is organised as follows: Sect.~\ref{sec:data} describes the data, measurements and methodology, Sect.~\ref{sec:anal} the analysis and results, Sect.~\ref{sec:disc} discusses our results and Sect.~\ref{sec:conc} presents our conclusions.

\section{SN sample and their properties}
\label{sec:data}

Throughout this series of papers, we used a historical sample of nearby SNe that has both spectra and photometry in various bands. The SN~Ia sample after cuts (see Sect.~\ref{sec:EWmeas}) is composed of 981 objects (with 3230 spectra). The sample contains only a few peculiar objects, such as Iax (8), 91bg-like (2), 91T-like (1) and super-Chandra (3). We will compare the narrow lines measurements from intervening material such as \naid\ with SN and host properties. A list of those properties and their median is provided in Tab.~\ref{tab:props}, whereas in App.~\ref{ap:longtable} we provide the full set of parameters for all SNe~Ia in this study. 

\begin{table}
\tiny
\centering
\caption{SN and environmental properties used in this study.}
\vspace*{-2mm}
\label{tab:props}
\renewcommand{\arraystretch}{1.2}
\begin{tabular}
{C{2.5cm}|C{1.0cm}C{1.0cm}C{1.0cm}C{1.5cm}}
\hline
\hline
\textbf{Property} & \textbf{Nr SNe~Ia}&  \textbf{MED} & \textbf{MAD} & \textbf{Source} \\ 
\hline
 \multicolumn{5} {c} {\textbf{Narrow line of \naid} (Sect.~\ref{sec:EWmeas})}  \\
\hline
EW(\AA) &  514 & $0.278$ & 0.493  & \citetalias{GG24} \\
VEL(\kms) & 327 & $4.0$ & 83.9 & \citetalias{GG24} \\
\hline
 \multicolumn{5} {c} {\textbf{Environmental properties} (Sect.~\ref{sec:envs})}  \\
\hline
$\overline{\Delta\alpha}$ & 834 & 0.203 & 0.107 & \citetalias{GG25} \\
$\log$ SFR$^L$(M$_{\sun}$yr$^{-1}$) & 359 & $-3.15$ & 1.42 & \citetalias{GG25} \\
$\log$ sSFR$^L$(yr$^{-1}$) & 359 & $-11.81$ & 0.27 & \citetalias{GG25} \\
$t_{\mathrm{age}}^L$(Gyr) & 359 & $1.51$ & 1.28 & \citetalias{GG25} \\
$A_V^L$ & 359 & $2.58$ & 0.72 & \citetalias{GG25} \\
$\log$ M$_*^L$(M$_{\odot}$) & 359 & $7.96$ & 0.72 & \citetalias{GG25} \\
$\log$ M$_*^G$(M$_{\odot}$) & 873 & $10.84$ & 0.70 & \citetalias{GG25} \\
\hline
 \multicolumn{5} {c} {\textbf{Photometric properties} (Sect.~\ref{sec:phot})}  \\
\hline
$s$ & 427 & 0.956 & 0.097 & This work$^{\ast}$ \\
$\mathcal{C}$ & 425 & 0.075 & 0.110 & This work$^{\ast}$ \\
$s_{BV}$  & 404 & 0.929 & 0.107 & This work$^{\dagger}$\\
$EBV$  & 404 & 0.208 & 0.100 & This work$^{\dagger}$ \\
$R_V$  & 404 & 3.285 & 1.300 & This work$^{\dagger}$ \\
$dBV_{60}$(mag\,d$^{-1}$) & 177 & $-0.011$ & 0.003 & This work$^{\star}$ \\
$BV_{60}$  & 177 & 0.839 & 0.152 & This work$^{\star}$ \\
\hline
 \multicolumn{5} {c} {\textbf{Spectral properties} (Sect.~\ref{sec:spec})}  \\
\hline
$v_{max}$(\kms)  & 274 & $-10937$ & 739 & This work \\
$v_{grad}$(\kmsd)  & 144 & 73.8 & 35.7 & This work \\
$v_{neb}$(\kms)  & 44 & $505$ & 1165 & This work \\
\hline
 \multicolumn{5} {c} {\textbf{Hubble residuals} (Sect.~\ref{sec:HR})}  \\
\hline
HR & 285 & $-0.010$ & 0.097 & This work$^\clubsuit$ \\
HR$_{L}$ & 182 & $-0.002$ & 0.104 & This work$^\clubsuit$\\ 
HR$_{G}$ & 259 & $0.000$ & 0.101 & This work$^\clubsuit$ \\
\hline
\end{tabular}
\vspace*{-2mm}
\tablefoot{Property, number of SNe with finite EW and property, median (MED), median absolute deviation (MAD) and source.\\
$\ast$ Using \textsc{SiFTO}\\
$\dagger$ Using \textsc{SNooPy}\\
$\star$ Following \citet{Forster13}\\
$\clubsuit$ Using \citet{Brout22}}
\end{table}

\subsection{Narrow line measurements}
\label{sec:EWmeas}

We developed a robust automatic tool to measure the equivalent width (EW in \AA) and velocity shift (VEL in km$/$s) of the narrow lines, as explained in \citetalias{GG24}. Since we found no evolution with time, when several spectra at different epochs exist for a given SN, we stack the flux-to-continuum ratios of each spectrum and do a bootstrap analysis. It is important to note that for low-resolution spectra, the P-Cygni profile of the fast-moving ejecta may interfere with the profile of the slow-moving narrow line, thus altering the EW measurement and leading to erroneous conclusions. By applying a continuum slope cut around the line, we make sure to use spectra for which this interference is minimal. Together with a signal-to-noise cut around the line, the initial sample of 1079 SNe~Ia (with 3534 spectra) is reduced to 981 objects (with 3230 spectra). Our automated technique integrates a fixed region that, in the case of noise or emission lines, can result in negative EW values. For the velocity measurement, we only consider cases with |EW|$>0.3$\AA\, to ensure there is enough absorption to calculate the wavelength of the minimum. The rest-frame zero velocity is taken with respect to the galaxy recessional velocity obtained from NED\footnote{The NASA/IPAC Extragalactic Database: \url{https://ned.ipac.caltech.edu/}.}, as in \citet{Maguire13}. Although internal rotations are not taken into account in this way, we expect that these effects are mostly washed out with large statistics.

\subsection{Environmental parameters}
\label{sec:envs}

In \citetalias{GG25} we introduced the environmental parameters calculated for our SN sample. By taking available multi-band images of the SN host galaxies from UV to IR, we do both: i) global photometry using Kron apertures and ii) local photometry taking apertures of 0.5 kpc in radius. The photometry is then fitted to composite stellar populations with a dust attenuation law to extract global and local parameters such as the stellar mass $M_*^L(M_{\odot})$, the stellar age $t_{\mathrm{age}}^L$(Gyr), the star formation rate SFR$^L$($M_{\odot}$yr$^{-1}$) and specific star formation rate sSFR$^L$(yr$^{-1}$), the dust attenuation $A_V^L$, among others. We focus here on these five local parameters, which are the ones showing the strongest relations to the narrow lines \citepalias{GG25}, as well as on the offset of the SN from the host centre normalised by the semi-major axis, $\overline{\Delta\alpha}$. We also include the global mass $M_*^G(M_{\odot})$ because of its relevance for cosmology (see Sect.~\ref{sec:HR}).

\subsection{Photometric properties}
\label{sec:phot}

\begin{figure}
\centering
\includegraphics[width=\columnwidth]{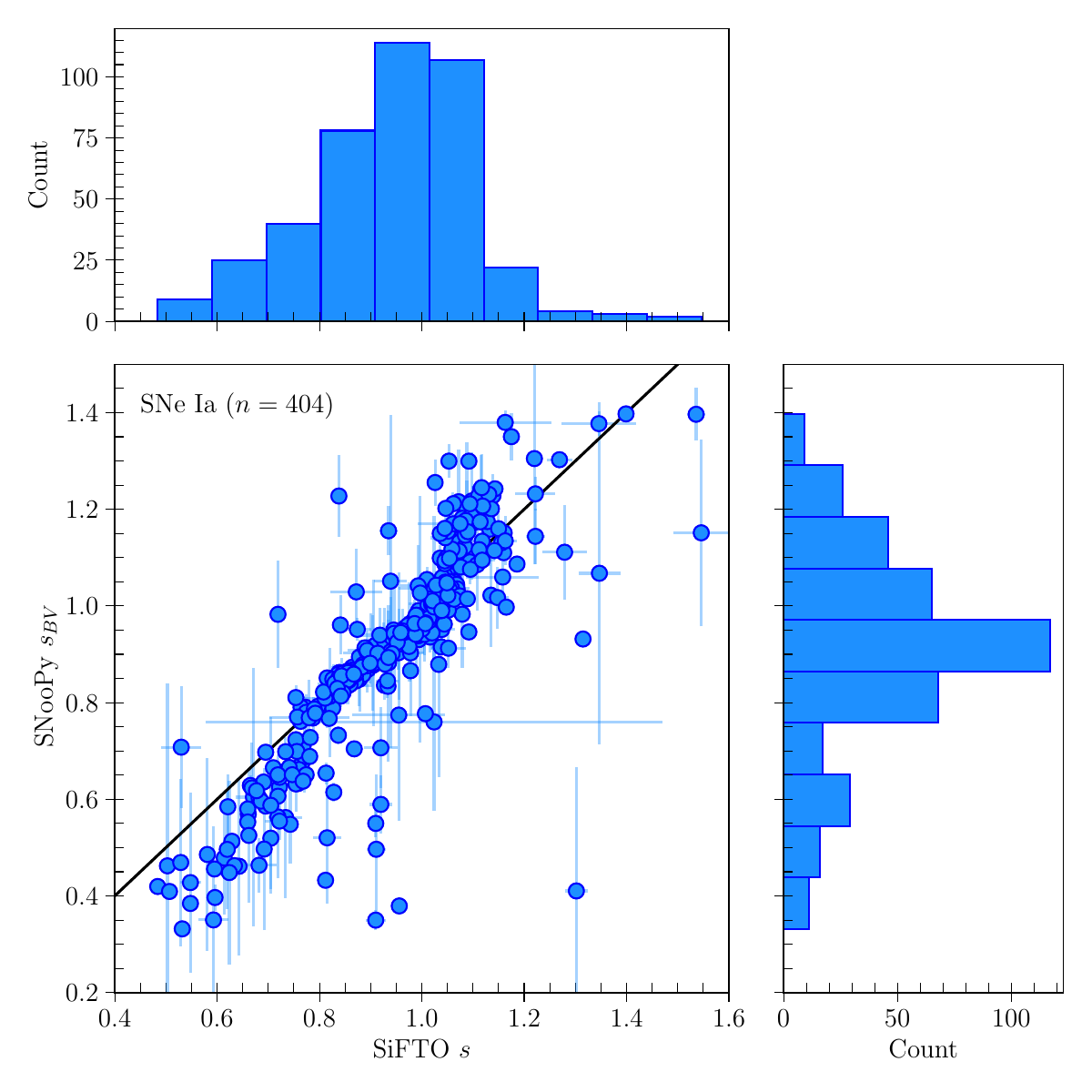}
\vspace*{-7mm}
\caption{\textsc{SNooPy} $s_{BV}$ vs \textsc{SiFTO} $s$ light-curve width parameters for SNe~Ia in our sample with their corresponding histograms in the side panels. }
\label{fig:stretch}
\end{figure}

\begin{figure}
\centering
\includegraphics[width=\columnwidth]{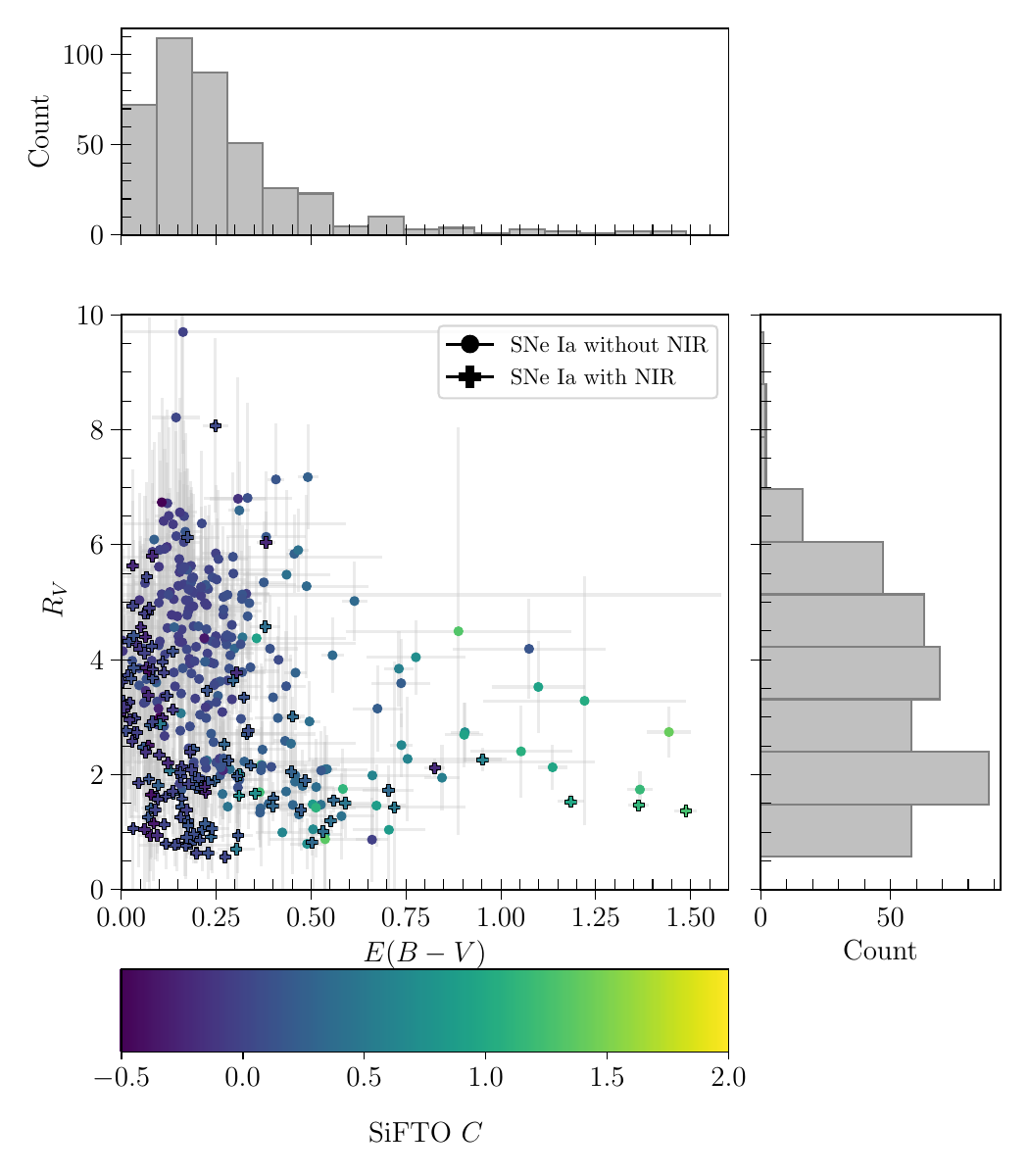}
\vspace*{-7mm}
\caption{\textsc{SNooPy} $R_V$ vs $EBV$ parameters for SNe~Ia in our sample with their corresponding histograms in the panels. The point colours indicate the SiFTO colour $\mathcal{C}$. Crosses indicate SNe with NIR used in the fits.}
\label{fig:RvEBV}
\end{figure}

In this paper, we will compare the EW and VEL of the narrow lines with SN properties. The main photometric properties are the light-curve width,  usually parametrised via the stretch parameter $s$, and the observed $B-V$ colour $\mathcal{C}$ at maximum. These are obtained through multi-band light-curve fits with the \textsc{SiFTO} software \citep{Conley08}.  With the \textsc{SNooPy} fitter \citep{Burns14}, we also obtain the colour-stretch parameter, $s_{BV}$, which is related to the width of the colour-curve, along with the dust colour excess, $EBV\equiv E(B-V)$ and the dust extinction law parametrised by the total-to-selective extinction ratio, $R_V=A_V/E(B-V)$. \textsc{SNooPy}, as opposed to \textsc{SiFTO}, assumes an intrinsic colour that depends on $s_{BV}$ from which the dust extinction and its wavelength dependence can be extracted assuming a reddening law parametrisation \citep[][CCM]{Cardelli89,ODonnell94}. Near-infrared (NIR) photometry, when available, greatly helps anchor the reddening curve. A distribution of these parameters for our sample is shown in Figures~\ref{fig:stretch} and ~\ref{fig:RvEBV}. Individual example fits are shown in App.~\ref{ap:details_phot}. As can be seen, the $R_V$ spans a wide range of values beyond the Milky Way (MW) observations with which the CCM law was obtained \citep[$2.6<R_V<5.6$,][]{Fitzpatrick99}. Beyond $R_V\sim5$, the reddening curve in the optical and NIR changes minimally, so that very high $R_V$ values are basically consistent with lower values, and this is well reflected in the larger error bars. Importantly, SNe with little extinction are more difficult to fit and have very uncertain $R_V$ estimates, stressing the importance of the prior (see App.~\ref{ap:details_phot}).   

We also investigate the photometric properties at later phases, between 35 and 85 days past maximum. This period is known for having a homogeneous linear decline in colour, also known as the Lira law \citep{Phillips99}, and is possibly a better telltale of the true LoS extinction than using colours at maximum. Furthermore, \citet{Forster13} found a possible trend between the rate of colour decline and the EW of sodium that is worth exploring further. We thus use late-time $B$ and $V$ photometry, corrected for MW extinction using $EBV$ maps from \citet{Schlafly11} and doing $K$-corrections using spectral templates \citep[see][]{Hsiao07,Forster13}, to calculate two parameters: the Lira-law $B-V$ slope $dBV_{60} \equiv d(B-V)/dt|_{60d}$ in mag$/$day, and the corresponding late-time colour $BV_{60} \equiv (B-V)_{60d}$. 

We thus have seven photometric parameters: $s$, $\mathcal{C}$, $s_{BV}$, $EBV$, $R_V$, $dBV_{60}$ and $BV_{60}$. Three of those should, in principle, be only related to the explosion, i.e. they are intrinsic properties of the SNe and reflect their inner evolution: $s$ and $s_{BV}$ represent the main evolution and are closely related to each other (see Fig.~\ref{fig:stretch}), as well as $dBV_{60}$, which describes the late evolution. On the other hand, $EBV$ and $R_V$ are purely \emph{extrinsic} parameters (assuming the intrinsic colours are well modelled) coming from dust extinction. Finally, the observed colours at maximum, $\mathcal{C}$, and at late times, $BV_{60}$, are a combination of intrinsic and extrinsic factors. 

\subsection{Spectral properties}
\label{sec:spec}

\begin{figure}
\centering
\includegraphics[width=\columnwidth]{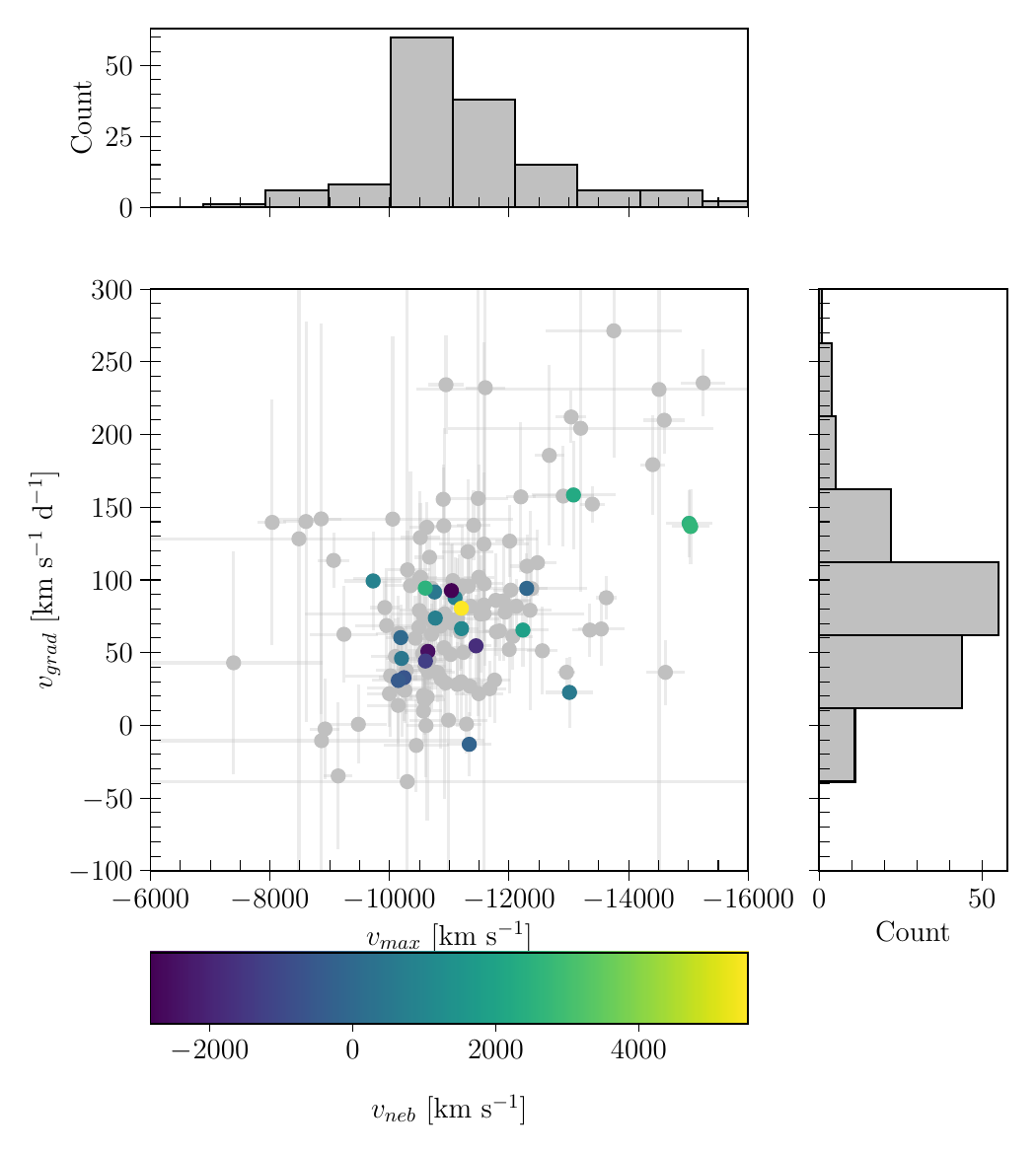}
\vspace*{-7mm}
\caption{Velocity gradient vs velocity at maximum for \ion{Si}{ii} (6355\AA) and their corresponding histograms. The point colours indicate the nebular velocity when available.} 
\label{fig:vel}
\end{figure}

A key spectroscopic property that describes the variability of SNe~Ia is the \ion{Si}{ii} (6355\AA) expansion velocity near maximum \citep{Benetti05}. Moreover, as we peer into deeper layers of slower-moving ejecta with time, the velocity gradient is also an important characteristic of the SN evolution \citep{Wang09}. We use our own semi-automatic code to measure the velocities of \ion{Si}{ii}. We define the continuum of the line with the same blue-ward and red-ward wavelength ranges of \citet{Folatelli13}. We then calculate the minimum of the smoothed line profile inside these bounds to obtain the wavelength shift and the expansion velocity. This is done for all available spectra of a given SN. We require at least one spectrum within $\pm3$ days of $B$-band maximum light to calculate the $v_{max}$(\kms) parameter from a linear interpolation. For the velocity decline rate, we estimate an expansion velocity similarly at 20 days after maximum from spectra $\pm5$ around that date, and use: $v_{grad} = (v_{max}-v_{20})/20$ \citep{Blondin12}. Fig.~\ref{fig:vel} shows these parameters for our sample. Some individual examples and more details are given in App.~\ref{ap:details_spec}.

We also investigate a very different spectroscopic regime when the SN has reached the nebular phase. At these times ($\gtrsim100$~d), we are observing the innermost part of the ejecta and the shifts of forbidden lines synthesized near the centre of the explosion such as [\ion{Ni}{ii}] $\lambda$7378\AA\, and [\ion{Fe}{ii}] $\lambda$7378\AA\, 7155\AA\, are useful to infer the explosion asymmetry \citep[e.g][]{Maeda10}. We use the same spectral literature sample as in \citetalias{GG24} to look for nebular spectra, finding 194 spectra for 64 SNe~Ia. We measure their nebular velocity, $v_{neb}$, following the procedure described by \citet{Maeda11}. When both nebular emission lines [\ion{Ni}{ii}] $\lambda$7378\AA\, and [\ion{Fe}{ii}] $\lambda$7378\AA\, 7155\AA, are detected, we determine the line shifts by fitting each feature with a Gaussian profile. The $v_{neb}$ is then defined as the mean velocity derived from the two lines. In cases where only one of the two lines is detected, $v_{neb}$ is measured using the available line alone. Many of the analysed SNe already had measurements that agree with our estimates \citep{Maeda11, Silverman12, Blondin12}.

\subsection{Hubble residuals}
\label{sec:HR}

SNe~Ia have been extensively used as distance indicators for cosmology \citep[e.g][]{Riess22,Popovic25}. As imperfect standard candles, their luminosity needs to be corrected for at least two parameters, the light-curve width and the colour. A third correction relating to the environment of the host galaxy, often represented by the host stellar mass \citep[e.g][]{Sullivan10,Lampeitl10}, accounts for additional variations not captured in the first two corrections. This third correction, also known as the "mass-step", has gathered significant attention in recent years, questioning its origin: extrinsic due to dust extinction, intrinsic due to differing SN properties, or a combination of both \citep[e.g.][]{Brout21,GG21,Duarte23,Wiseman23}.

In this work, we also study the dependence of the Hubble residuals on the narrow lines of intervening material. For this, we use the SN parameters given in \citet{Brout22}, notably the light-curve width $x_1$ and the colour at maximum $c$ obtained with the \textsc{SALT} fitter \citep{Guy07,Guy10,Brout22SALT}. We assume a fixed $\Lambda$CDM cosmology with $H_0=73$ \kms\,Mpc$^{-1}$, $\Omega_M=0.30$ and $\Omega_{\Lambda}=0.70$, so that:
$$    \mathrm{HR} = \mu_{\mathrm{SN}}(M_B,\alpha,\beta,\Delta_{\mathrm{host}}) - \mu_{\mathrm{cosmo}}(z,H_0,\Omega_M,\Omega_{\Lambda}), 
$$
\noindent with:
\vspace*{-2mm}
$$    \mu_{\mathrm{cosmo}} = 5\log(d_L/10\mathrm{pc}),
$$
\noindent where $d_L$ is the cosmology-dependent luminosity distance, and

$$    \mu_{\mathrm{SN}} = m_B -M_B + \alpha x_1-\beta c +\delta_{\mathrm{host}},
$$
\vspace*{-2mm}

\noindent with the mass-step given by:
\vspace*{-2mm}
$$
\delta_{\mathrm{host}}=\begin{cases}
			\Delta_{\mathrm{host}}, & \text{if $M<M_{\mathrm{step}}$}\\
            -\Delta_{\mathrm{host}}, & \text{if $M>M_{\mathrm{step}}$}
		 \end{cases}
$$
\noindent with $M_{\mathrm{step}}$ taken as the median of the sample.

The parameters $M_B$, $\alpha$, $\beta$ and $\Delta_{\mathrm{host}}$ are free parameters of the cosmological fit.  We perform fits and obtain Hubble residuals (HR) for three cases: one without a mass-step, another with a local mass-step obtained here from the stellar mass within the $r=0.5$ kpc aperture, and a third one with a global mass-step from the stellar mass of the full Kron aperture (see Sect.~\ref{sec:envs}). The HR of these three fits are called: HR, HR$_L$ and HR$_G$, and they do not include bias corrections. 
However, we emphasise that in this study, our objective is not to obtain accurate cosmological fits and standardisation parameters, but to explore relative differences of HR for various SN narrow line abundances. Tab.~\ref{tab:cosmo} summarises the fitted parameters. More details can be found in App.~\ref{ap:details_hr}.

\begin{table}
\centering
\tiny
\caption{Standardisation parameters from cosmological fits.}
\vspace*{-2mm}
\label{tab:cosmo}
\renewcommand{\arraystretch}{1.4}
\begin{tabular}{c|ccc}
\hline
\hline
\backslashbox[1.3cm]{Par}{Fit} & \makecell{No $\delta_{\mathrm{host}}$\\(HR)} & \makecell{Local  $\delta_{\mathrm{host}}$\\(HR$_L$)} & \makecell{Global  $\delta_{\mathrm{host}}$\\(HR$_G$)} \\
\hline
 N$_{\mathrm{SN}}$ & 359 & 195 & 277 \\
  $M_B$ & $-19.215(10)$ & $-19.184(24)$ & $-19.186(17)$ \\
 $\alpha$ & 0.155(09) & 0.166(14) & 0.166(11) \\
 $\beta$ & 2.690(112)  &  2.834(185) & 2.695(176)\\
 $\Delta_{\mathrm{host}}$ & -- & $-0.067(31)$ & $-0.073(25)$ \\
 $\sigma_{int}$ & 0.105(12) & 0.122(21) & 0.103(28)\\
 RMS  & 0.234 & 0.245 & 0.211\\
\hline
\end{tabular}
\end{table}

\section{Analysis and Results}
\label{sec:anal}

In this section, we analyse the narrow line properties, particularly the EW of \naid, compared to the different intrinsic and extrinsic properties presented in the previous section. We employ mainly statistical means to compare the property distributions. The Kolmogorov-Smirnov (KS) test \citep{ks-test1,ks-test2} permits testing the hypothesis that two different distributions of a single quantity arise from the same parent population. The Fasano-Franceschini (FF) test \citep{Fasano87}, on the other hand, extends the KS test to the comparison of two samples with multiple properties. In both cases, a p-value lower than 0.05 rejects the hypothesis that both samples come from the same parent population. In \citetalias{GG24} and \citetalias{GG25}, we also consider a bootstrap analysis that ensures that both compared distributions are consistent in redshift, while allowing the dividing value between both samples to be swept between 40-60\% of the full distribution to obtain a final median p-value with a corresponding probability ($P_{MC})$ of being lower than 0.05. In Sect.~\ref{sec:simple}, we study the narrow line EW distributions when divided according to single properties, in Sect.~\ref{sec:double} we repeat the study, ensuring that the environments of the compared SN properties are more consistent, while in Sect.~\ref{sec:clust} we obtain SN groups with a clustering analysis and investigate their narrow lines.

\subsection{\naid\ EW distributions}
\label{sec:simple}

\begin{table}
\tiny
\centering
\caption{KS tests of two EW distributions divided according to a single property.}
\vspace*{-2mm}
\label{tab:KSsimp}
\renewcommand{\arraystretch}{1.2}
\begin{tabular}
{C{1.5cm}|C{1.55cm}C{1.55cm}C{1.2cm}C{1.2cm}}
\hline
\hline
\textbf{Property} & $<\mathbf{EW}_{\mathrm{hi}}>$ & $<\mathbf{EW}_{\mathrm{lo}}>$ & \textbf{KS}$^{\star}$ &$\mathbf{P_{MC}}$$^{\dagger}$ \\ 
\hline
 \multicolumn{5} {c} {\textbf{Environmental properties} }  \\
\hline
$\overline{\Delta\alpha}$ & 0.08$\pm$ 0.29 & 0.71$\pm$0.66  & 3.80e-15 & 100(100) \\
$\log$ SFR$^L$ & 0.84$\pm$0.69 & 0.09$\pm$0.26 & 7.56e-10 &  100(100)\\
$\log$ sSFR$^L$ & 0.73$\pm$0.69 & 0.12$\pm$0.28 & 8.33e-7 & 100(98) \\
$t_{\mathrm{age}}^L$ & 0.13$\pm$0.31 & 0.61$\pm$0.63 & 1.29e-4 & 92(67) \\
$A_V^L$ & 0.59$\pm$0.63 & 0.19$\pm$0.38 &  4.67e-2& 41(9)\\
$\log$ M$_*^L$ & 0.73$\pm$0.73 & 0.13$\pm$0.28 & 2.10e-6 & 100(97) \\
$\log$ M$_*^G$ & 0.28$\pm$0.51 & 0.39$\pm$0.54 & 2.96e-1 & 10(2) \\
\hline
 \multicolumn{5} {c} {\textbf{Photometric properties} }  \\
\hline
$s$ & 0.37$\pm$0.45 & 0.18$\pm$0.46 & 3.20e-3 &  88(50)\\
$\mathcal{C}$ & 0.61$\pm$0.67 & 0.08$\pm$0.33 & 1.19e-5 & 100(97) \\
$s_{BV}$  & 0.41$\pm$0.45 & 0.13$\pm$0.48 & 1.56e-4 & 90(63)\\
$EBV$  & 0.80$\pm$0.66 & 0.01$\pm$0.26 & 1.74e-11 & 100(98) \\
$R_V$  & 0.21$\pm$0.43 & 0.37$\pm$0.52 & 2.69e-1 &  19(4)\\
$dBV_{60}$ & 0.17$\pm$0.37 & $0.15\pm0.36$ & 6.21e-1 & 6(0) \\
$BV_{60}$  & 0.54$\pm$0.51 & $-0.06\pm0.18$ & 8.05e-8 & 100(98) \\
\hline
 \multicolumn{5} {c} {\textbf{Spectral properties}}  \\
\hline
$v_{max}$  & 0.37$\pm$0.54 & 0.30$\pm$0.45 & 3.22e-1 & 9(1) \\
$v_{grad}$  & 0.24$\pm$0.49 & 0.26$\pm$0.32 & 1.37e-1 & 15(2)\\
$v_{neb}$  & 0.63$\pm$0.68 & $-0.05\pm0.16$ & 1.38e-2 & 34(6) \\
\hline
 \multicolumn{5} {c} {\textbf{Hubble residuals}}  \\
\hline
HR & 0.15$\pm$0.39 & 0.13$\pm$0.38 & 9.09e-1 &  5(0)\\
HR$_{L}$ & 0.30$\pm$0.35 & 0.12$\pm$0.28 & 3.25e-1 & 9(1) \\ 
HR$_{G}$ & 0.24$\pm$0.40 & 0.12$\pm$0.37 & 8.71e-1 & 6(0) \\
\hline
\end{tabular}
\vspace*{-2mm}
\tablefoot{Property, median and MAD EW(\AA) of \naid\ for the upper/lower half of the property sample, raw KS test and bootstrap probability (\%) for the p-value$<$0.05 ($<$0.01 in parentheses). \\
$\star$ Raw KS test between EW distributions divided by their median\\ 
$\dagger$ Bootstrap probability (\%) of p-value$<$0.05 ($<$0.01) between $z$-matched EW distributions divided by sweeping values within 40-60\% }
\end{table}

We start by highlighting that the strong trends found in \citetalias{GG25} of \naid\ EW with local host properties are maintained with SNe~Ia: there is a stronger abundance of sodium in SNe~Ia located in central regions and local environments that are more star-forming, younger, more massive and that have a higher dust attenuation. These findings, summarised in Tab.~\ref{tab:KSsimp}, are obtained from individual KS tests that divide the SN sample into two groups, according to the median of the property considered (see Tab.~\ref{tab:props} for the median values) or values swept around the median for the bootstrap probability. All the p-values are well below 0.05, and their probabilities are above 80\% (except for $A_V^L$). This is also consistent with the local H$\alpha$ from the host relating to \naid\ EW for SNe~Ia, as found in \citet{Anderson15Ia}. For the global mass, we do not find that the two EW distributions are different from each other, confirming that \naid\ is a tracer of local gas abundance (see \citetalias{GG25}) and that local and global properties do not follow a one-to-one relation.

Regarding the SN properties, we see that the light-curve width, parametrised both by $s$ and $s_{BV}$, shows a strong relation with the EW of \naid\ ($P_{MC}\sim90$\%). This is expected as high-stretch SNe~Ia are known to occur in younger and more star-forming regions \citep[e.g.,][]{Hamuy95,Hamuy96}, so there is a stronger abundance of gas and dust around them. The observed colour $\mathcal{C}$ is highly correlated with the EW ($p\sim1e^{-5}$, $P_{MC}=100$\%; see also \citealt{Anderson15Ia}), confirming that the sodium abundance is a good tracer of dust extinction, as has been extensively shown previously \citep[e.g.][]{Munari97,Murga15}. This is even more evident when we use the colour excess, $EBV$ ($p\sim 1e^{-11}$, $P_{MC}=100$\%), which is in principle exclusively due to extinction without intrinsic colour variations. Similarly, the observed colour in the Lira law regime, i.e. 60 days after maximum, is indeed a better tracer of the extinction than at maximum light ($p\sim 1e^{-8}$, $P_{MC}=100$\%), as the intrinsic variations among SNe~Ia have decreased by that time.

\begin{figure}
\centering
\includegraphics[width=0.9\columnwidth]{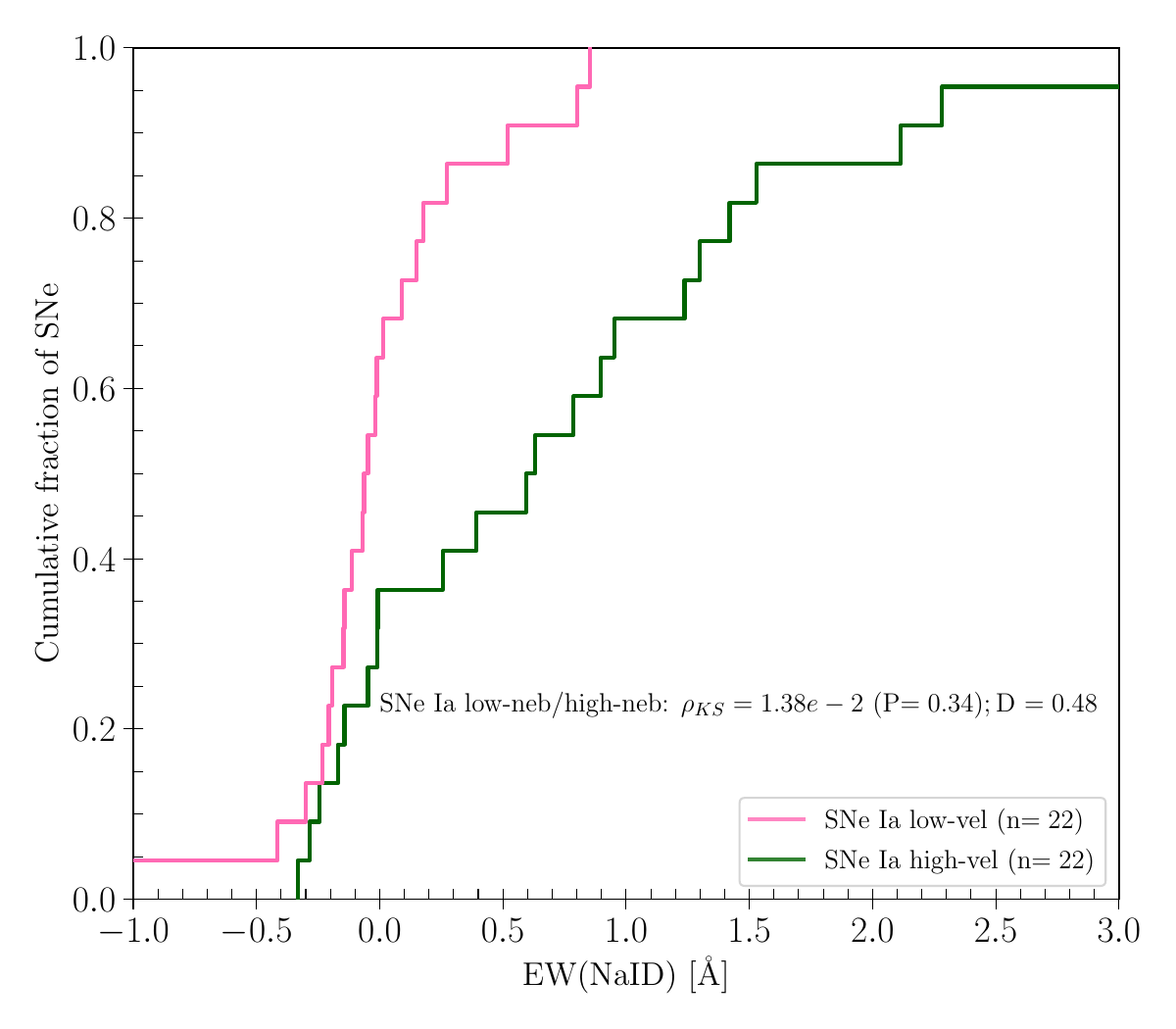}
\vspace*{-3mm}
\caption{Cumulative distributions of \naid\ EW divided according to the median of the nebular velocity $v_{neb}$. 
}
\label{fig:vneb}
\end{figure}

Interestingly, splitting the sample according to the nebular velocity shows strong differences in EW (see Fig.~\ref{fig:vneb}) with a p-value of 0.014. The low $v_{neb}$ sample has a median EW of $-0.05$\AA\ compared to 0.63\AA\ for the high $v_{neb}$ sample. The bootstrap probability is rather small (34\%), but this is due to the small sample size of 44 SNe. Such a trend was already found in \citet{Forster12} and indicates a relation between intrinsic SN properties, i.e. the asymmetry of the explosion, and its surrounding material. 

Other parameters do not show significant p-values and high bootstrap probabilities, but have interesting differences in their median EW values. SNe~Ia with lower $R_V$ have higher EW (0.40 vs 0.24\AA), a result that agrees with previous findings of higher extinction for steeper reddening laws \citep[e.g.,][]{Mandel11,Burns14}. The Hubble residuals corrected for mass-step (both local and global) have median EW values that are larger for higher HR compared to lower HR (0.30 vs 0.12\AA). This is an interesting result that we explore further in Sect.~\ref{sec:mass-step}. 

\subsection{\naid\ EW distributions in similar environments}\label{sec:double}

\begin{table}
\tiny
\centering
\caption{Median (and median absolute deviation) for environmental properties of two subsamples: young (YG) and old (OLD).}
\vspace*{-2mm}
\label{tab:2pops}
\renewcommand{\arraystretch}{1.4}
\begin{tabular}
{p{0.8cm}|p{0.25cm}p{1.15cm}p{1.15cm}p{0.8cm}p{0.8cm}p{0.8cm}}
\hline
\hline
Sub-sample & Nr & <sSFR$^L$> & <SFR$^L$> & <$t_{\mathrm{age}}^L$> & <$A_V^L$> & <$M_*^L$>\\
\hline
OLD & 222 & $-12.1(0.1)$ & $-4.3(1.2)$ & 2.8(2.0) & 2.6(0.9) & 7.7(0.9) \\
\hline
YG & 137 & $-10.5(0.6)$ & $-1.5(0.7)$ & 0.3(0.3) & 2.6(0.5) & 8.3(0.5) \\
\hline
\end{tabular}
\tablefoot{The values are logarithmic for sSFR, SFR and $M_*$.}
\end{table}

\begin{table*}
\tiny
\centering
\caption{KS tests of two EW distributions divided according to a single property for two environmental subsamples: old (left), young (right).}
\vspace*{-2mm}
\label{tab:KSdoub}
\renewcommand{\arraystretch}{1.3}
\begin{tabular}
{C{1.5cm}|C{1.52cm}C{1.52cm}C{1.04cm}C{0.9cm}C{1.05cm}||C{1.52cm}C{1.52cm}C{1.04cm}C{0.9cm}C{1.05cm}}
\hline
\hline
\textbf{Subsample} & \multicolumn{5} {c||} {\textbf{OLD}} & \multicolumn{5} {c} {\textbf{YOUNG}}  \\
\hline
\textbf{Property} & $<\mathbf{EW}_{\mathrm{hi}}>$ & 
$<\mathbf{EW}_{\mathrm{lo}}>$ & \textbf{KS}$^{\star}$ &$\mathbf{P_{MC}}$$^{\dagger}$ & \textbf{FF}$_{env}$$^{\triangle}$ & $<\mathbf{EW}_{\mathrm{hi}}>$ & $<\mathbf{EW}_{\mathrm{lo}}>$ & \textbf{KS}$^{\star}$ &$\mathbf{P_{MC}}$$^{\dagger}$ & \textbf{FF}$_{env}$$^{\triangle}$ \\ 
\hline
 \multicolumn{11} {c} {\textbf{Environmental properties} }  \\
\hline
$\overline{\Delta\alpha}$ & 0.01$\pm$0.20 & 0.59$\pm$0.58 & 6.97e-4 & 93(65) & 1.45e-4 & 0.25$\pm$0.39 & 0.85$\pm$0.73 & 2.13e-2 & 13(2) & 4.60e-2\\ 
\hline
 \multicolumn{11} {c} {\textbf{Photometric properties} }  \\
\hline
$s$ & 0.24$\pm$0.34 & 0.07$\pm$0.45 & 3.84e-2 &  32(6) & 3.46e-3 & 1.20$\pm$0.80 & 0.61$\pm$0.66 & 2.88e-1 &   11(2) & 5.90e-1\\
$\mathcal{C}$ & 0.39$\pm$0.59 & 0.05$\pm$0.18 & 8.10e-3 & 72(26)  & 5.16e-1 & 1.05$\pm$0.77 & 0.41$\pm$0.61 & 2.12e-1 & 11(2)  & 6.83e-1 \\
$s_{BV}$  & 0.39$\pm$0.49 & 0.01$\pm$0.30 & 4.80e-3 & 81(31)  & 1.98e-1 & 1.05$\pm$0.94 & 0.91$\pm$0.67 & 8.66e-1 & 6(0) & 6.97e-1\\
$EBV$  & 0.51$\pm$0.65 & 0.01$\pm$0.15 & 4.10e-4 & 92(61)  & 4.23e-1 & 1.25$\pm$0.64 & 0.21$\pm$0.36 & 1.42e-3 & 70(22)  & 1.56e-1\\
$R_V$  & 0.24$\pm$0.36 & 0.09$\pm$0.49 & 3.95e-1 & 13(2)  & 1.04e-1& 0.25$\pm$0.40 & 1.42$\pm$0.69 & 1.59e-3 & 92(61)  & 6.86e-1\\
$dBV_{60}$ & 0.13$\pm$0.31 & 0.04$\pm$0.23 & 9.74e-1 & 2(0)  & 6.00e-3 & 0.45$\pm$0.57 & 0.84$\pm$0.74 & 6.33e-1 & 1(0) & 7.98e-3\\
$BV_{60}$ & 0.59$\pm$0.46 & $-0.12\pm$0.13 & 1.02e-4 &  99(95) & 7.61e-1 & 0.84$\pm$0.58 & 0.05$\pm$0.29 & 1.73e-2 & 22(4)  & 2.89e-1 \\ 
\hline
 \multicolumn{11} {c} {\textbf{Spectral properties}}  \\
\hline
$v_{max}$ & 0.27$\pm$0.42 & 0.01$\pm$0.29 & 2.58e-1 & 13(2) & 6.32e-1 & 0.26$\pm$0.49 & 1.30$\pm$0.56 & 8.39e-3 & 79(30)  & 5.97e-1 \\
$v_{grad}$  & 0.18$\pm$0.42 & 0.06$\pm$0.21 & 3.48e-1 &  4(0) & 5.51e-1 & 0.85$\pm$0.98 & 1.25$\pm$0.99 & 6.48e-1 &  7(0) & 4.89e-1\\
$v_{neb}$  & 0.59$\pm$0.36 & $-$0.01$\pm$0.14 & 3.52e-2 & 12(2)  & 2.82e-1 & 1.24$\pm$0.98 & $-0.23\pm$0.18 & 6.37e-3 & 6(0) & 6.15e-1\\
\hline
 \multicolumn{11} {c} {\textbf{Hubble residuals}}  \\
\hline
HR & 0.15$\pm$0.28 & 0.05$\pm$0.17 & 1.40e-1 &  18(3) & 1.02e-1 & 0.45$\pm$0.59 & 0.41$\pm$0.64 & 9.31e-1 & 2(0)  & 2.21e-3 \\
HR$_{L}$& 0.24$\pm$0.35 & 0.01$\pm$0.13 & 4.94e-3 & 78(29)  & 5.02e-1 & 0.45$\pm$0.59 & 0.41$\pm$0.64 & 9.76e-1 & 2(0) & 4.10e-3 \\ 
HR$_{G}$ & 0.19$\pm$0.31 & 0.01$\pm$0.13 & 1.22e-2 & 31(6)  & 1.72e-1 & 0.45$\pm$0.59 & 0.41$\pm$0.64 & 9.33e-1 & 2(0)  & 2.09e-2\\
\hline
\end{tabular}
\vspace*{-2mm}
\tablefoot{Property, median and MAD EW(\AA) of \naid\ for the upper/lower half of the property sample, raw KS test and bootstrap probability (\%) for the p-value$<$0.05 ($<$0.01 in parentheses).\\
$\star$ Raw KS test p-value between EW distributions divided by their median\\ 
$\dagger$ Bootstrap probability (\%) of p-value$<$0.05 ($<$0.01) between $z$-matched EW distributions divided by sweeping values within 40-60\% \\
$\triangle$ Multi-dimensional FF test p-value on the local environmental properties of the two EW samples}
\end{table*}

\begin{figure}
\centering
\includegraphics[width=0.9\columnwidth]{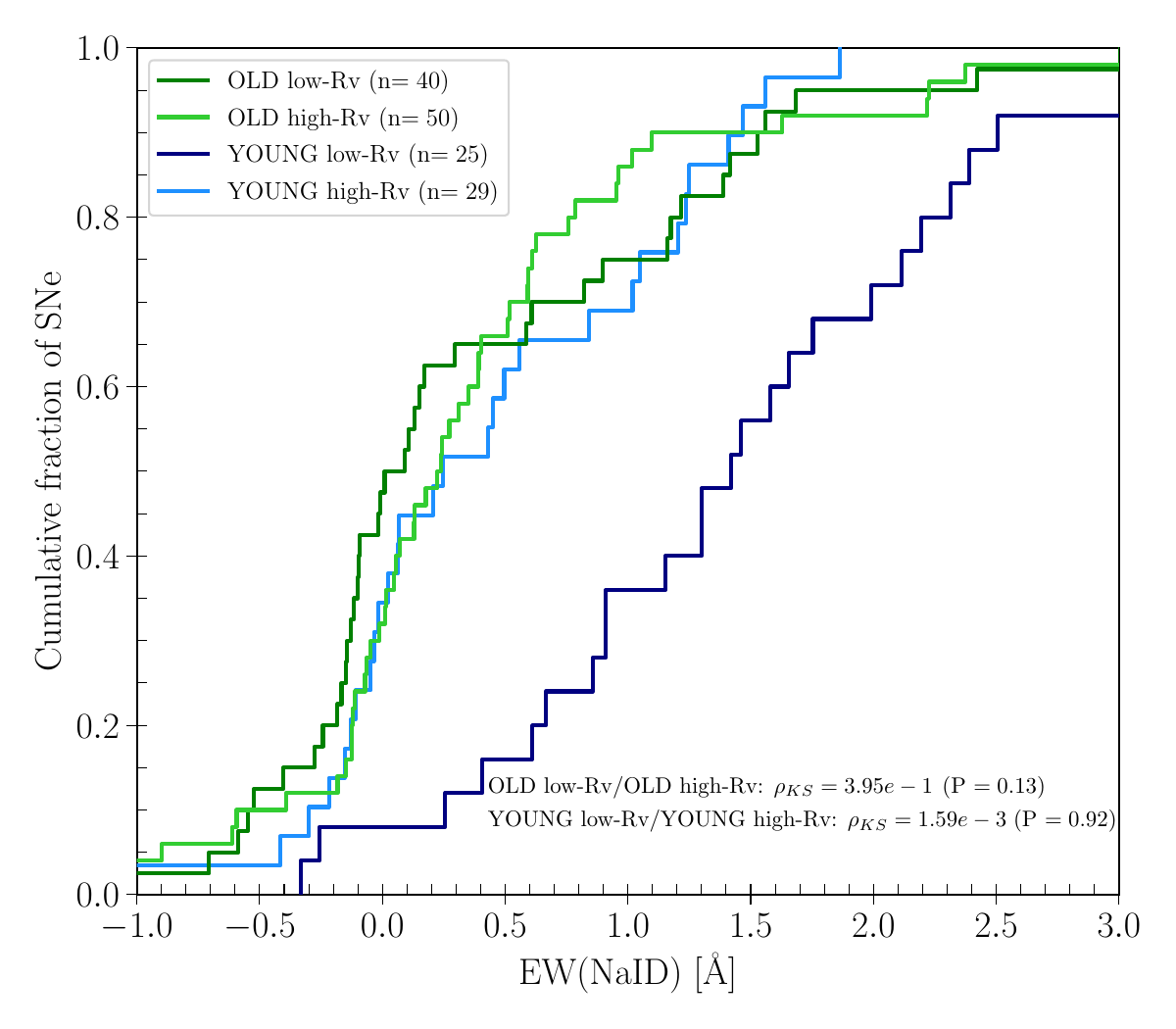}
\includegraphics[width=0.9\columnwidth]{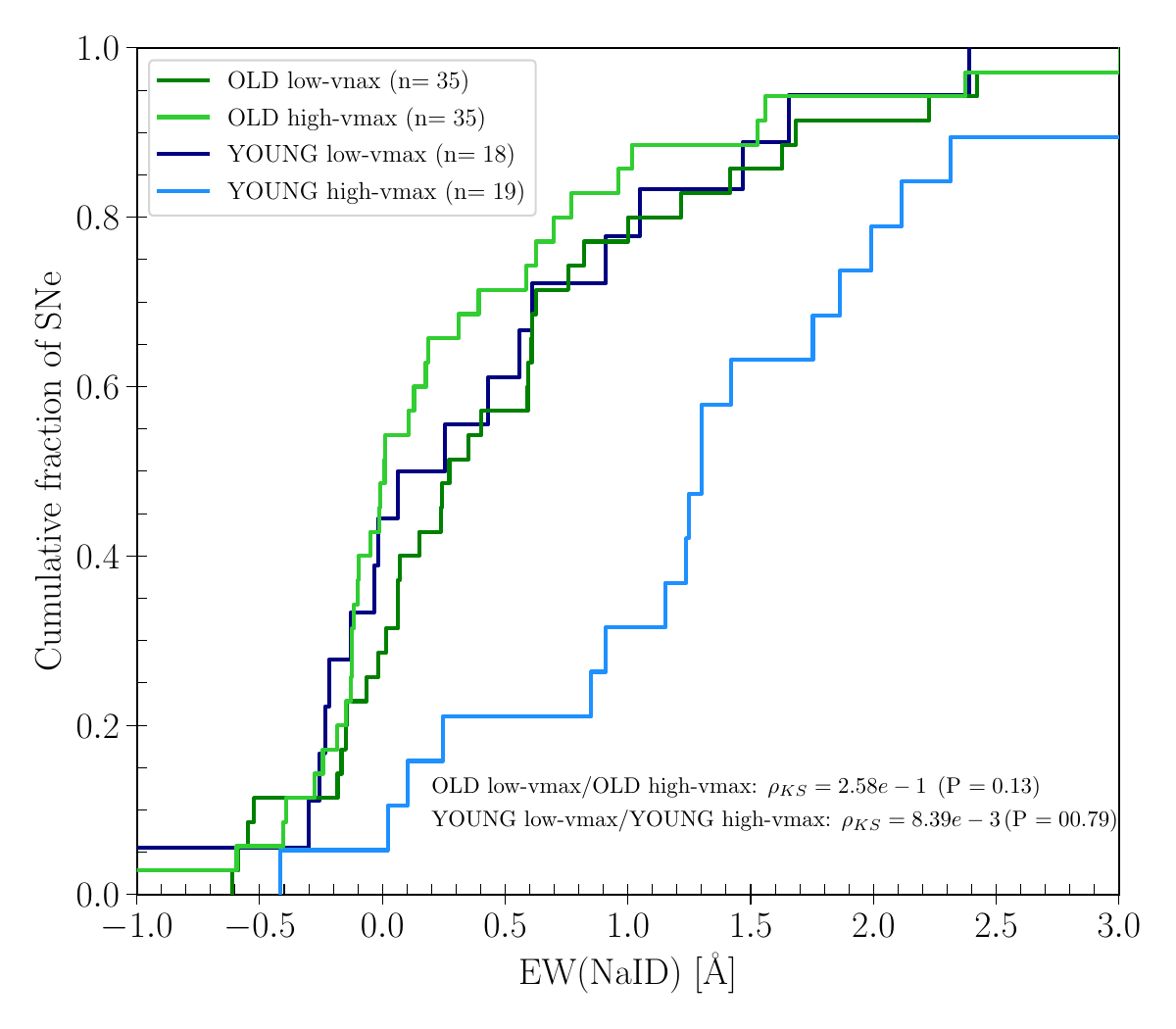}
\vspace*{-2mm}
\caption{Cumulative distributions of \naid\ EW divided by the median of the $R_V$ (top) and the $v_{max}$ (bottom) for the old and young subsamples.
}
\label{fig:Rv,vmax}
\end{figure}
\vspace*{-2mm}

In the previous section, we found important differences in \naid\ EW absorption according to various SN properties. However, this may be entirely driven by the dependence of SN properties (e.g., stretch) on the host. In order to exclude these effects, we attempt to divide the sample according to the environments. 

We use the multi-dimensional version of the KS test, the FF test, which combines multiple properties of two samples to find a p-value giving the probability that these two samples do not arise from the same parent population. We explore a division that separates the best two samples according \emph{only} to their local environmental properties of SFR, sSFR, age, dust attenuation and stellar mass. This experiment (see App.~\ref{ap:FFenv} for details) results in a best division of local $\log($sSFR$)=-11.3$ that splits the sample into two populations of SNe~Ia with strongly differing environments. We call these the "old" and "young" subsamples (see Tab.~\ref{tab:2pops}). Two environmental populations of SNe~Ia have been suggested early in the literature with a prompt (younger stellar populations) and a delayed (older stellar populations) channel \citep{Mannucci05,Sullivan06}. More recently, less-biased samples of SNe~Ia find evidence for bimodal distributions, particularly in the light-curve width parameter \citep{Nicolas21,Ginolin25}, which strongly correlates with environment. In our case, the division is rather arbitrary but provides a means to study the EW distributions in more detail. 

After separating the sample into two subpopulations with significantly different environments, we repeat the individual KS tests of the EW distributions divided according to SN properties for each of these two environmental subsamples. The results are summarised in Tab.~\ref{tab:KSdoub} for the old (left) and young (right) components, respectively. Despite the division into two environmental populations, there might be non-discrete variations with environment within each subsample considered. We therefore add a column that provides the FF test of the two property populations arising from the same distribution based on all local properties.  

For the old population, we find that the EW distributions divided according to the light-curve width ($s$ and $s_{BV}$) show significant differences (p$<0.04$) which probably are still entirely driven by environmental differences, as the FF p-values are also significant (at least for $s$ with p$<0.05$). On the other hand, the observed colour at maximum ($\mathcal{C}$) and during the Lira regime ($BV_{60}$), as well as the colour excess $EBV$, have significantly different EW distributions with simultaneously consistent environments (FF p$>0.4$), indicating that there are important dust variations in the SN environments at scales smaller than those probed by our local estimates of $r=$0.5 kpc. Moreover, the nebular velocity stands out again, as SNe with high-redshifted velocities show stronger sodium absorption. Finally, the Hubble residuals corrected for mass-step show relevant differences in EW, particularly for the local mass-step (p$=0.013$), with higher HR values (fainter after other corrections) having larger EWs.

For the young component, the division according to light-curve width results in consistent EW distributions. However, the $EBV$ and $R_V$ parameters have significant p-values revealing that SNe~Ia, which are more extincted and with lower $R_V$, possess a much larger abundance of sodium ($\gtrsim1.2$\AA) than their counterparts ($\lesssim0.25$\AA). These large differences in EW are also seen for populations divided according to the ejecta velocity (SNe with higher  $v_{max}$\footnote{In the table the <EW$_{\mathrm{lo}}$> for $v_{max}$ refers to more negative velocities.} have larger EW) and the nebular velocity (higher receding velocities have larger EW). We show the cumulative distributions of EW divided according to $R_V$ and to $v_{max}$ for the old and young populations in Fig.~\ref{fig:Rv,vmax}. A relation of high-velocity SNe~Ia with redder colours and lower $R_V$ values had already been found \citep{Wang09,Foley11}. In App.~\ref{ap:SFRcorr} we show that even after correcting the EW for the local SFR, the relations with SN properties are maintained.  

The striking result that intrinsic SN properties, such as the ejecta or nebular velocities, correlate with properties of the intervening material, such as the \naid EW, beyond local environmental characteristics, suggests that either sodium is a proxy for another physical property related to the explosion, or that the SN interacts with very nearby material in the LoS. In asymmetric explosions, the central asymmetry, given by the nebular velocity, goes in the opposite direction of the outer material, shown by the photospheric velocities \citep{Maeda10,Maguire18}. In the framework of the DDet scenario, the first outer helium shell detonation propagates around the surface to the opposite side, leading to an off-centre second detonation \citep{Townsley19,Li21}. A core detonation observed from the initial ignition side will have ejecta moving at higher photospheric velocities, but preferentially redshifted receding nebular velocities.

\subsection{Cluster analysis}\label{sec:clust}

In the previous section, we divided the SN~Ia sample into two groups according to their local environments. However, this division was rather arbitrary and did not include SN properties, which may reveal clearer differences between real intrinsic populations. Furthermore, the posterior division into SN properties was done by taking the median of each property, meaning that each new division could include different SNe. Instead of finding artificial one-dimensional divisions that are not realistic of true sub-populations with possible overlapping properties, we use here a different approach consisting of clustering algorithms that consider all features, both intrinsic and environmental. 

We use Gaussian Mixture Models \citep[GMM,][]{Mclachlan00} that model the data through a probabilistic combination of Gaussians representing each subpopulation (see App.~\ref{ap:clust} for details). In this clustering algorithm, the number of clusters is predetermined by the user. We use local environmental variables (SFR$^L$ and $t_{\mathrm{age}}^L$) and SN properties ($s$, $R_V$ and $v_{max}$) as input features to the GMM. Since we aim to study the differences in EW for various clusters, we omit in the clustering any narrow line information as well as any intrinsic property directly related to dust, such as colour (e.g., $\mathcal{C}$, $EBV$, $BV_{60}$). Additional variables were also tested with similar results. However, as we increase the number of input features, the amount of SNe with available properties diminishes. These five properties result in 106 SNe~Ia. The GMM algorithm provides the probability of each SN belonging to each cluster, and the largest probability provides the cluster membership. Additionally, we consider the uncertainties in the input parameters by doing a Monte Carlo that shifts the variables according to their covariance matrix and recalculates the probability of belonging to each cluster at each iteration. The final probability is the median of all simulations.

\begin{figure}
\centering
\includegraphics[width=0.98\columnwidth]{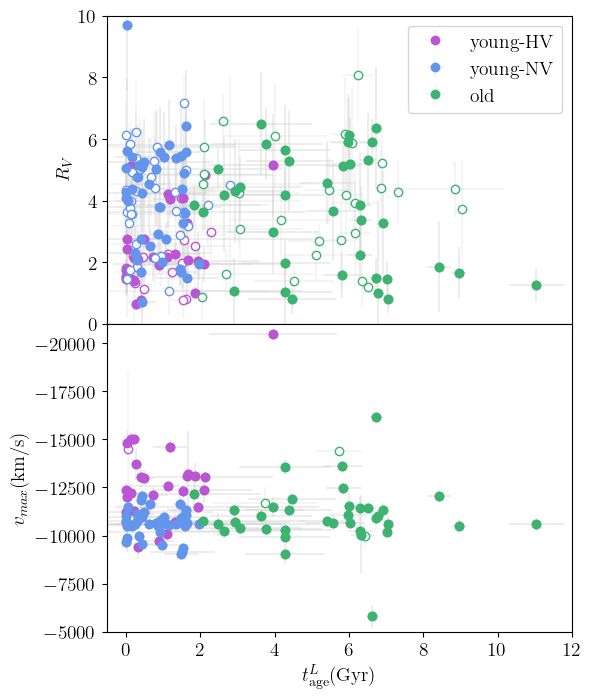}
\vspace*{-4mm} 
\caption{Reddening law $R_V$ and ejecta velocity $v_{max}$ vs local age $t_{\mathrm{age}}^L$ for three clusters obtained through GMM: an "old" component (green), a "young-NV" (blue) and a "young-HV" population (purple). Filled circles represent the initial SN sample, and open circles are SNe with one imputed variable. 
}
\label{fig:3cluster}
\end{figure}

\begin{table}
\tiny
\centering
\caption{Median properties for three SN clusters: "old", "young-NV", and "young-HV"}
\vspace*{-2mm}
\label{tab:3clust}
\renewcommand{\arraystretch}{1.3}
\begin{tabular}
{C{1.5cm}|C{1.92cm}|C{1.92cm}|C{1.92cm}}
\hline
\hline
& \multicolumn{3}{|c}{Cluster} \\
\hline
\textbf{Property} & \textbf{OLD} & \textbf{YOUNG-NV}  & \textbf{YOUNG-HV} \\
\hline
Nr & 44 & 29 & 33 \\
EW &$-0.06\pm0.30$ & $0.35\pm0.47$ & $1.30\pm0.69$ \\
VEL & $24\pm207$ &  $111\pm248$ &  $-67\pm221$  \\
\hline
$\overline{\Delta\alpha}$ & 0.36$\pm$0.10 &  0.16$\pm$0.06 & 0.12$\pm$0.04 \\
\textbf{SFR}\boldsymbol{$^\mathrm{L}$} & $-5.08\pm$1.02 &  $-2.47\pm$0.68 &  $-2.41\pm0.87$  \\
\boldsymbol{$t_{\mathrm{age}}^L$} & 4.43$\pm$2.45 &  0.82$\pm$0.61  & $0.49\pm0.47$  \\
\hline
\boldsymbol{$s$} & 0.87$\pm$0.09 & 0.91$\pm$0.15 & 0.98$\pm$0.05 \\
$\mathcal{C}$ & $0.07\pm0.09$  & $0.06\pm0.11$ &  $0.24\pm0.24$  \\
$s_{BV}$  & 0.89$\pm$0.09 &  $0.81\pm0.18$  & 0.96$\pm$0.05\\
$EBV$  & $0.19\pm0.09$  & $0.21\pm0.10$  & $0.37\pm0.20$  \\
\boldsymbol{$R_V$}  & 3.81$\pm$1.56&  4.76$\pm$0.69 &  2.13$\pm$0.61 \\
$dBV_{60}$ & $-0.011\pm0.001$ &  $-0.011\pm0.003$ &  $-0.013\pm0.003$ \\
$BV_{60}$ &$0.85\pm0.11$  &$0.82\pm0.12$ & $1.04\pm0.21$ \\ 
\hline
\boldsymbol{$v_{max}$} & $-10718\pm481$ & $-10630\pm409$ &  $-12296\pm1008$ \\
$v_{grad}$ & $66.5\pm21.3$  & $67.7\pm31.1$  & $87.4\pm44.7$  \\
$v_{neb}$ & $-182\pm673$ &  $-482\pm1554$  & $1427\pm974$  \\
\hline
HR & 0.08$\pm$0.14 & $-0.01\pm0.09$  & $-0.02\pm0.12$ \\
HR$_{L}$ & 0.06$\pm$0.15 & $0.04\pm0.09$ & $0.00\pm0.12$ \\ 
HR$_{G}$ & 0.05$\pm$0.15 & $0.01\pm0.06$ & $-0.04\pm0.13$ \\
\hline
\end{tabular}
\tablefoot{Property, median and MAD for three clusters: old, young-NV and young-HV (see Tab.~\ref{tab:3clustIMP} for the imputed sample). The five input properties used in the clustering are highlighted in bold.}
\end{table}

We first fix the number of clusters to two and find results similar to the previous section: environmental properties drive the separation between a young and an old component of SNe~Ia, while SN properties play a very minor role in this separation. When using three clusters, the intrinsic properties emerge: there is one group of old SNe~Ia, whereas the young group is now subdivided into two subpopulations, one with preferentially low $R_V$ and high $v_{max}$ and another with higher $R_V$ and lower $v_{max}$ (see Fig.~\ref{fig:3cluster}). Differences in other input features are less evident. Using four groups results roughly in the same clusters, plus a very small additional group of a few outlier members. We therefore stop at three clusters. The Akaike information criterion \citep[AIC, e.g.][]{AIC}, which calculates the fit quality penalising additional parameters from an increasing number of clusters, confirms the three-member clustering. 

Tab.~\ref{tab:3clust} shows the median properties of all three groups that we dub "old", "young-NV and "young-HV", names that refer to the age and the ejecta velocity of the clusters\footnote{The old population fully corresponds with the division of Sect.~\ref{sec:2pops}, i.e. they all have $\log$ sSFR$<11.3$, whereas the two young channels have over 60\% of their environments with $\log$ sSFR$>11.3$.}. Besides the input parameters used for clustering, we also show the median of other SN properties, as well as the narrow line EW and velocities. To augment the sample, we also perform an imputation of missing data: SNe with four out of the five input parameters are imputed, and the new set is used to obtain a membership class. The extended imputed sample consists of 192 objects, and the statistics for this sample are also shown in Tab.~\ref{tab:3clustIMP}. The imputation is used only for labelling into a given cluster, but the statistics come from real measured data (see App.~\ref{ap:clust} for details). 

\begin{figure}
\centering
\includegraphics[width=\columnwidth]{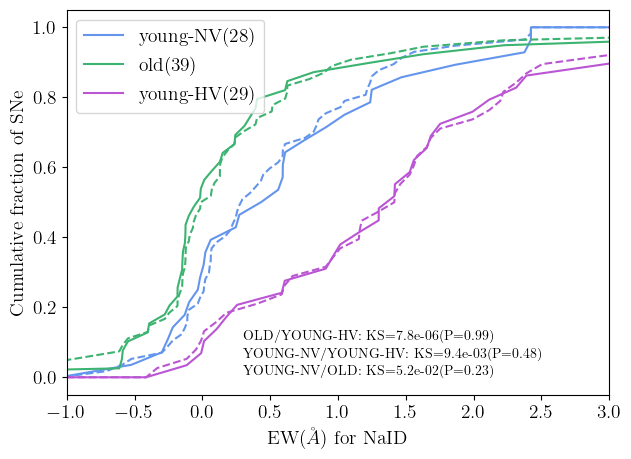}
\vspace*{-6mm}
\caption{Cumulative \naid\ EW distribution for three clusters obtained through GMM: an "old" component (green), a "young-NV" (blue) and a "young-HV" population (purple) for the normal (solid lines) and the imputed sample (dashed lines).  
}
\label{fig:3clust-cum}
\end{figure}

As expected, the old population has the lowest EW of \naid\, mostly due to the different environment. We confirm that the two young populations have significant differences in EW that correlate with other extinction-related parameters such as the observed colour, $EBV$ and $R_V$, but also to other properties such as the ejecta velocity, the velocity gradient and the nebular velocity. Compared to the young-NV population, the young-HV sample has much stronger EW ($1.30$ vs $0.35$\AA) as also shown in Fig.~\ref{fig:3clust-cum}, larger extinction ($EBV=0.37$ vs 0.21), steeper reddening curves ($R_V=2.12$ vs 4.76), higher ejecta velocities ($v_{max}=-12296$ vs $-10630$ \kms), larger ejecta velocity gradients ($v_{grad}=87.4$ vs 67.7 \kmsd) and higher redshifted nebular velocities ($v_{neb}=1427$ vs $-482$ \kms). These findings are confirmed for the increased imputed sample.

\begin{figure*}
\centering
\includegraphics[width=0.83\textwidth]{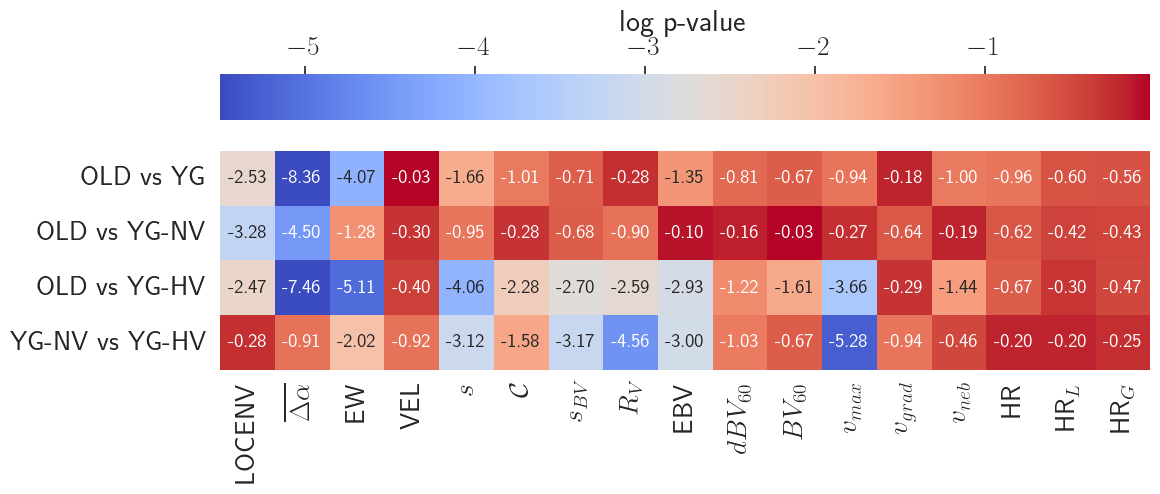}
\vspace*{-2mm}
\caption{KS test p-values for different properties divided according to three clusters: "old" (OLD), "young-NV" (YG-NV) and "young-HV" (YG-HV. The first row joins both young groups into a single one (YG). "LOCENV" refers to a combination of local environmental properties (sSFR$^L$, SFR$^L$, $t_{\mathrm{age}}^L$, $A_V^L$ and M$_*^L$) through a FF test. For reference, a p-value of 0.05 corresponds to $\log$(p-val)$=-1.30$. 
}
\label{fig:3clustKSmatrix}
\end{figure*}

In Fig.~\ref{fig:3clustKSmatrix}, we show the KS tests among the three clusters for various properties. These three clusters confirm that normal-velocity (NV) SNe~Ia can be found in two groups (old and young-NV), whereas high-velocity (HV) objects form exclusively a different group (young-HV), as found in \citet{Li21}. It is worth noting that \cite{Burns14} also find evidence for a two-Gaussian mixture model for the fitted $R_V$ parameter. Using their findings as an alternative prior to $R_V$ strengthens our results.

\subsection{Other narrow lines}\label{sec:other-lines}
\begin{figure*}
\centering
\includegraphics[width=0.32\textwidth]{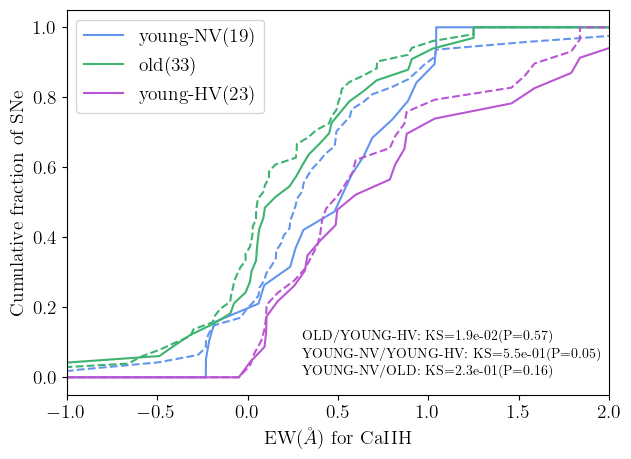}
\includegraphics[width=0.32\textwidth]{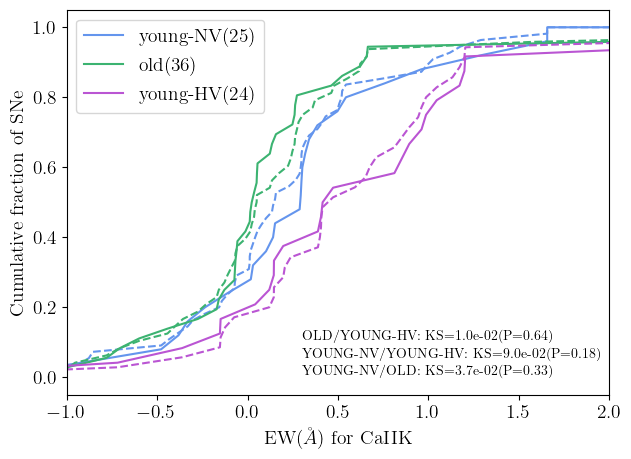}
\includegraphics[width=0.32\textwidth]{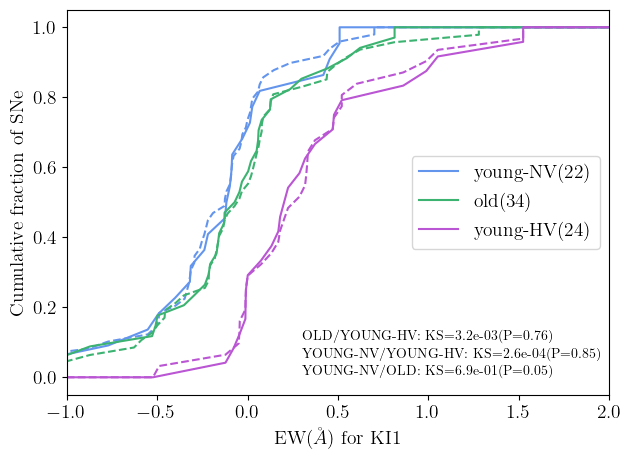}
\vspace*{-3mm}
\caption{Cumulative \naid\ EW distribution for three GMM clusters, old (green), young-NV (blue) and young-HV (purple), for \caii\ H (left), \caii\ K (middle) and \ki\ 1 for the normal (solid lines) and the imputed sample (dashed lines).  
}
\label{fig:3clust-cum-other}
\end{figure*}

We show here the distribution of EW for other narrow lines, divided according to the three GMM clusters previously found. We confirm that the young-HV group presents stronger absorption also for \caii\ H\&K and for \ki\ 1, although the differences are less significant than for sodium (see Fig.~\ref{fig:3clust-cum-other} and Tab.~\ref{tab:3clust-lines}). In spite of this, some lines, notably the \ion{K}{i} 2 and the DIB lines, do not show any relevant difference among the groups. This could possibly be important for physical interpretations; however, these lines are also weaker and more difficult to measure with low-resolution spectra, and the sample is even smaller in these cases.

\begin{table}
\tiny
\centering
\caption{Median EW line properties for three SN clusters}
\vspace*{-2mm}
\label{tab:3clust-lines}
\renewcommand{\arraystretch}{1.3}
\begin{tabular}
{C{1.5cm}|C{1.92cm}|C{1.92cm}|C{1.92cm}}
\hline
\hline
\textbf{EW} & \textbf{OLD} & \textbf{YOUNG-NV}  & \textbf{YOUNG-HV} \\
\hline
\naid\ &$-0.06\pm0.30$ & $0.35\pm0.47$ & $1.30\pm0.69$ \\
\caii\ H &$0.10\pm0.20$ & $0.48\pm0.40$ & $0.50\pm0.38$ \\
\caii\ K &$0.02\pm0.18$ & $0.29\pm0.26$ & $0.42\pm0.44$ \\
\ki\ 1 &$-0.10\pm0.19$ & $-0.12\pm0.19$ & $0.19\pm0.20$ \\
\ki\ 2 &$-0.02\pm0.17$ & $-0.05\pm0.18$ & $0.00\pm0.25$ \\
DIB 5780 &$0.04\pm0.17$ & $0.07\pm0.15$ & $0.11\pm0.12$ \\
DIB 4428 &$0.13\pm0.22$ & $0.07\pm0.20$ & $0.21\pm0.43$ \\
DIB 6283 &$-0.07\pm0.09$ & $-0.11\pm0.20$ & $0.01\pm0.16$ \\
\hline
\end{tabular}
\tablefoot{Property, median and MAD for three clusters (see Tab.~\ref{tab:3clust-lines-IMP} for the imputed sample).}
\end{table}

\section{Discussion}
\label{sec:disc}

In this section, we study the results in greater detail, in light of SN~Ia models and findings of the literature. 

\subsection{Two populations}
\label{sec:2pops}

\begin{figure}
\centering
\includegraphics[width=0.96\columnwidth]{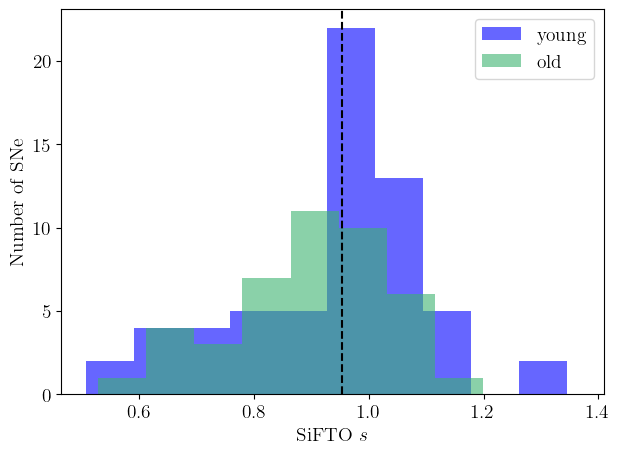}
\vspace*{-2mm}
\caption{\textsc{SiFTO} stretch distribution for the old and young (both NV and HV) components obtained with the GMM clustering. The vertical dashed line is the division value of $s=0.954$ ($x_1=-0.48$) from \citet{Ginolin25}, which is close to the median of the entire sample ($<s>=0.959$).
}
\label{fig:3clust_ST}
\end{figure}

In Sect.~\ref{sec:double}, we divided the sample into two groups according to their local environment and, in Sect.~\ref{sec:clust}, we found that a clustering analysis of two and three clusters recovers these two environmental populations. Such a division is not surprising and has been previously suggested in SN~Ia studies as a prompt component and a delayed component \citep{Mannucci05,Sullivan06,Greggio08}. More recently, there has been evidence for two populations divided by the light-curve width parameter, which strongly correlates with the environment \citep{Nicolas21,Ginolin25}. In Fig.~\ref{fig:3clust_ST}, we show the stretch distribution for the "old" and "young" populations (both NV and HV) found with our clustering. Although these two stretch populations are distinct (KS p-value of 0.02, see Fig.~\ref{fig:3clustKSmatrix}) and the separation found in \cite{Ginolin25} seems to isolate the majority of them, there is a clear overlap between the two. Fast-decliners (low stretch) SNe~Ia can also be found in the old population, and slow-decliners (high stretch) in the young population. 

\begin{figure*}
\centering
\includegraphics[width=0.85\textwidth]{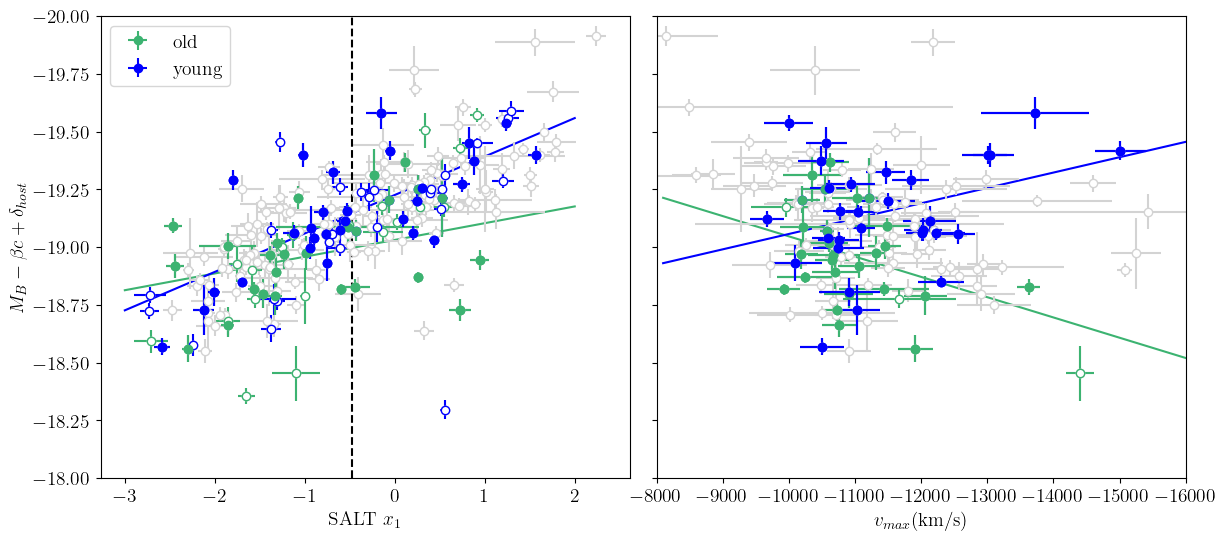}
\vspace*{-3mm}
\caption{Absolute magnitude corrected for colour (and mass-step) versus  \textsc{SALT} $x_1$ (left) and \ion{Si}{ii} ejecta velocity (right) for the old (green) and young (blue) components obtained with GMM clustering (open circles for the imputed sample). Grey open dots are objects without cluster membership. 
}
\label{fig:3clust_mag-x1-vel}
\end{figure*}

More than a simple separation in light-curve width or in age of the entire SN~Ia population, \citet{Ginolin25} show that the brightness-stretch relation for their two stretch samples is significantly different (see their Fig.~7), which has profound implications in the powering mechanism of these two populations and therefore in their explosions and progenitors. In the left panel of  Fig.~\ref{fig:3clust_mag-x1-vel}, we show the SN magnitudes (corrected for distance and for the colour and mass-step calibrations) as a function of stretch ($x_1$) for our two clusters. Fitting a line using \textsc{linmix} \citep{Kelly07}, we obtain different slopes for the two groups, although within the uncertainties ($-0.072\pm0.043$ and $0.166\pm0.033$ for the old and young populations, respectively, or a 1.7$\sigma$ difference). 

\citet[][see their Fig.~11]{Polin19} also argue that the SN~Ia population could have two channels based on the relation of their brightness and ejecta velocity. We find slightly stronger differences in this parameter space (see right panel of Fig.~\ref{fig:3clust_mag-x1-vel}) with the slopes of old and young samples differing in sign, ($-8.88\pm5.99$) and ($6.70\pm4.27$) $\times10^{-5}$ \kms\ for the old and young populations, respectively, or a 2.1$\sigma$ difference. However, the fits are heavily dependent on a few higher-velocity objects. In the interpretation of \citet{Polin19}, there is a group of SNe~Ia with no strong dependence of the brightness on \ion{Si}{ii} velocity, which corresponds to \Ch\ explosion models, whereas the group with a brightness-velocity relation originates from \subCh\ DDet explosions for which increasing WD mass (between 0.9 and 1.2 $M_{\odot}$) results in brighter (between -17.5 and -20.0 magnitudes) and faster ejecta (between 9500 and 16000 \kms). If our relations were to be confirmed with more data, the young channel would agree better with the \subCh\ scenario. 

In Tab.~\ref{tab:KSdoub} we also see that the stretch within the old channel is still highly dependent on the environment (low FF$_{\mathrm{env}}$ of 3.5e-3), probably indicating that age may be a fundamental parameter responsible for the variability of the old SNe~Ia. The correlation between local age and stretch ($-0.41$, p-value of 0.005) confirms that younger SNe~Ia have higher light-curve widths, as expected in the typical brightness-stretch relation \citep{Phillips93}. The young population, on the other hand, shows much less dependence of the stretch on environment (FF$_{\mathrm{env}}$ of 5.9e-1 and correlation with age of $-0.28$, p-value of 0.04), indicating that another mechanism may be at hand. 

The age differences between both populations cannot be due to the cooling age of a WD, as this is inversely related to its mass, so that if the young population resulted from less massive \subCh\ WDs, they should be older. This would mean that the age is determined by a younger companion rather than the WD itself.  Although the normalised offset location was not used in the division nor the clustering, we find that both groups have significant differences: the old population has a median offset of 0.36 versus 0.16 for the young channel (KS test p-value of $4.4e-9$ with $P_{MC}$=100\%). This agrees with known age gradients due to the inside-out growth of galaxies \citep[e.g.,][]{Perez13}. The range of ages covered by the old population is larger than for the young one (see Tab.~\ref{tab:2pops}), and double-degenerate mergers can, in principle, explain such diversities better \citep{Maoz14}.

Regarding the sodium absorption, it is clear that the young population encounters more gaseous material in the LoS, as is expected in more star-forming regions. Indeed, there are clear differences between the two populations that remain related to the environment, e.g. both channels have stronger absorption closer to the centre of the hosts, although this seems again more important for the old population.

\subsection{Young asymmetric channel}\label{sec:asymm}

In our study, the so-called young component of SNe~Ia is further subdivided through clustering into two groups that seem to be related to explosion asymmetries through differences in the nebular velocity that also correlate with photospheric ejecta velocity and its gradient over time. This division separates SNe~Ia into the known spectral categories \citep{Wang09} of normal velocity (NV) and high-velocity (HV) objects, whereas the old component is composed primarily of only NV SNe (see Tab.~\ref{tab:3clust}). Such a subdivision of the NV group into two further groups has already been pointed out \citep{Kawabata20,Li21}. As this HV group preferentially happens for the young channel, which is more centrally located (see Tab.~\ref{tab:3clust}), the observed preference for HV closer to the centre of galaxies is confirmed \citep{Wang13,Pan20,Nugent23}. 

HV SNe~Ia have preferentially redshifted nebular velocities corresponding to asymmetric central explosions moving away from the observer, whereas NV SNe~Ia present more blueshifted nebular velocities moving towards the observer. These observations can be explained in the light of the DDet explosion mechanism, in which a \subCh\ WD  has an outer helium layer that detonates and creates a supersonic wave propagating on the surface and coinciding on the opposite side to trigger a secondary off-centre carbon-oxygen detonation \citep[e.g.,][]{Nomoto82,Hoeflich96,Woosley11,Shen14,Polin19}. \citet[][see their fig.~4]{Townsley19} demonstrate with multi-dimensional simulations that the \ion{Si}{II} velocity is more blueshifted when the first detonation is viewed face-on: depending on the viewing angle, the ejecta velocity can change from 9000 \kms\ when viewed from the opposite side to 17000 \kms\ when viewed from the same side. This arises because the core detonation is less curved and stronger as it reaches the surface \citep[see also multi-dimensional simulations,][]{Holas25}.

The sample of young-HV SNe~Ia has thus larger ejecta velocities and more redshifted nebular velocities than the young-NV one, in agreement with the \subCh\ DDet scenario. We also find that the young-HV objects have slightly higher stretch ($<s_{YG-HV}>=0.98$ vs $<s_{YG-NV}>=0.91$, see Tab.~\ref{tab:3clust}) and are more luminous (see right Fig.~\ref{fig:3clust_mag-x1-vel}). Sub-\Ch\ explosion models do show a relation between stretch, magnitude and viewing angle; however, it often goes in the opposite direction: surface He detonations viewed face-on are fainter and shorter-lived \citep{Townsley19,Collins25}. Nonetheless, in the edge-lit scenario, in which the core ignition is triggered between the core and the shell right after the He detonation \citep{Livne90}, the explosions with initial surface detonations on the side of the observer are brighter and longer-lived \citep{Gronow21}. However, this scenario would not explain the redshifted nebular velocities since both detonations happen on the same side. In general, a relation between $v_{max}$, stretch and brightness is expected in the \subCh\ scenarios, but the strength of this relation and its dependences on WD mass, viewing angle, and offset from the core \citep{Holas25} vary among studies. We caution that other asymmetric SN~Ia scenarios exist \citep[e.g.,][]{Pollin25}. Finally, it is also possible that there are two different populations in similar environments, i.e. that the young-NV and young-HV groups have distinct progenitor or explosion mechanisms.

\subsection{Nearby material}\label{sec:nearbydust}

The EW of the \naid absorption lines in SNe~Ia is related to the intervening material in the LoS. The old population has less star formation and cleaner environments, so they have lower EW than the young channel, as one would expect. On the other hand, the young-HV and young-NV groups have strong differences in sodium absorption that are seemingly not related to environments. This is accompanied by strong differences in observed colours and varying reddening laws: young-HV SNe~Ia are redder, they have steeper reddening curves and present stronger sodium absorption. Moreover, although the significance is lower, the young-HV sample has more blueshifted sodium lines, indicating gas movement towards the observer (see Tab.~\ref{tab:3clust}).

\begin{figure}
\centering
\includegraphics[width=0.96\columnwidth]{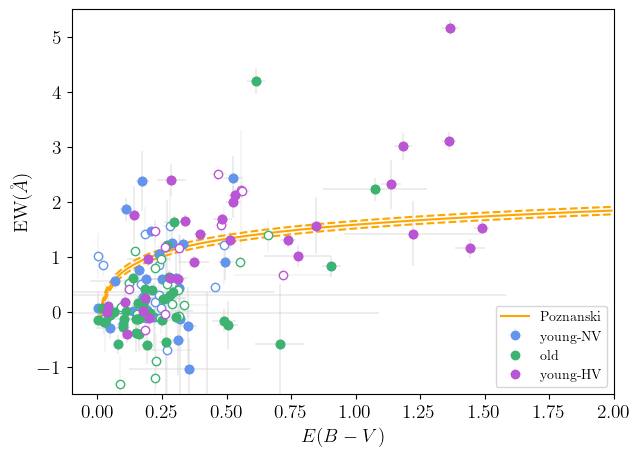}
\vspace*{-4mm}
\caption{\naid\ EW vs $E(B-V)$ for SNe~Ia divided according to the three clusters for the initial sample (filled circles) and the imputed sample (open circles). The Poznanski relation is shown in orange.  
}
\label{fig:3clust_EW-EBV}
\end{figure}

In Fig.~\ref{fig:3clust_EW-EBV}, we show the EW of sodium compared to the $EBV$ values from \textsc{SNooPy} for the three clusters found. For comparison, the relation obtained by \citet{Poznanski12} is shown in orange. Although there are caveats to using low-resolution spectra to estimate extinction \citep{Poznanski11} and there is a lot of scatter, it is interesting to see a higher fraction of young-HV SNe~Ia above the line (51\% are above and 38\% above 1$\sigma$ of their uncertainty) compared to the other two groups (the old group has 10\% above and 8\% above 1$\sigma$, and the young-NV has 21\% above and 11\% above 1$\sigma$). This indicates that there is an excess of sodium absorption for the young-HV group. \citet{Phillips13} found that indeed some SNe~Ia presented stronger sodium absorption than expected from a calibration of the MW, and that these excess objects had in fact blueshifted absorption. This agrees well with the sample of young-HV SNe, which are also more blueshifted on average. 

HV objects, belonging mostly to the young-HV sample, have been associated to redder colours \citep{Foley11}, lower $R_V$ values \citep{Wang09} and stronger sodium absorption \citep[e.g.,][]{Gall24}. Intrinsic colour differences not captured in the stretch-dependent intrinsic colours of \textsc{SNooPy} could affect the reddening estimates, as suggested by \citet{Polin19}; however, the accompanying strong differences in \naid\ absorption argue for real dust differences. The line-of-sight material responsible for these differences is 
more local than our $r=0.5$ kpc host apertures around the SNe, and more strikingly, is related to intrinsic explosion properties such as ejecta velocity and nebular shifts.  The most plausible explanation is that the SN interacts with material in its vicinity, with explosion asymmetries and viewing angle effects related to this interaction and its strength. 

At later times (30-90 days after maximum), \citet{Forster13} claim that sodium absorption differs according to the $B-V$ evolution. In this study, we only find mild indications for this (KS p-value of 0.09 between young-HV and young-NV), and it would seem that the diversity of this decay has a stronger relation to some property of the environment (see FF$_{env}$ in Tab.~\ref{tab:KSdoub}). It is also worth mentioning that \citet{Wang19} find late-time blue-excess (60-100 days past maximum) for HV objects, and they interpret this as evidence for light-echoes from very nearby material, in agreement with our findings.

Two main alternatives for the nearby material have been debated in the literature: CSM ejected from the progenitor system prior to explosion ($\lesssim1$pc from the SN), and nearby dust clouds from the ISM ($>$1 pc). The following evidence further argues for nearby material around SNe~Ia, and we highlight why the ISM interpretation is preferred: i) light-echoes in SNe~Ia are known to exist \citep[e.g.,][]{Wang08,Crotts08,Yang17}, yet only for a few SNe and the inferred distances for the dust are larger than 10 pc; ii) peculiar steep reddening curves (low $R_V$) in SNe~Ia, especially for HV objects as mentioned before, are best explained by normal dust \citep{Burns14,Marino15} rather than for a CSM-like material \citep{Goobar08}; iii) time-evolution in $EBV$ and $R_V$ \citep[e.g.,][]{Forster13} are well modeled by scattering of dust located at distances larger than $>1$ pc for most SNe~Ia \citep{Bulla18}; iv) time-varying narrow lines such as sodium or potassium occur only in some SNe~Ia with high-resolution spectra \citep{Patat07,Simon09,Graham15}, whereas statistical low-resolution sample do not show ubiquitous variation \citepalias{GG24}, this is consistent with geometrical effects from dust clouds in a patchy ISM affecting only some objects \citep[see also][]{Patat10,Maeda16}; v) excess blueshifts in narrow features of sodium \citep{Sternberg11,Clark21} associated to additional absorption  \citep{Maguire13,Phillips13} can be explained by mass-loss episodes prior to explosion in the CSM scenario, but they can also arise from acceleration of ISM dust clouds due to the SN radiation \citep{Hoang17}. 

Another line of evidence that argues for ISM is the lack of time-evolving polarisation in SNe~Ia (\citealt{Kawabata14,Patat15,Nagao18}, although see \citealt{Yang18}). Moreover, the increased continuum polarisation (and its bluer peak) occurs preferentially in arms of spiral galaxies and closer to the centre \citep{Zelaya17,Chu22}, arguing for increased IS dust. Indeed, one of the leading arguments against CSM is the location of these SNe~Ia that show signatures of nearby material: they are in central, star-forming, dusty regions, agreeing with an IS origin. We confirm that this is the case for the young-HV channel, which has more intervening material. But the young-NV channel has indeed almost the same environments and suffers from much less extinction. So, the signatures of nearby material go beyond environments, and there could be a coincidence of effects: the progenitors of the young dusty channel could eject CSM that is only viewed at certain angles. 

The standard processes capable of changing the column density of sodium are i) ionisation and recombination of the intervening gas, particularly efficient for very close material (CSM), and ii) geometrical effects, in which the SN photosphere gradually covers different fractions of material, which in this case can be farther away (ISM). The photo-ionisation of neutral sodium by the intense early UV SN radiation can cause an initial decrease of the column density (any sodium gas within 3 pc will be entirely ionised, see e.g. \citealt{Ferretti16}); later, the sodium recombines on a timescale that depends on the gas electron density. For the recombination to happen within weeks to be observed in the spectral evolution, a high electron density ($10^5-10^7$cm$^{-3}$), characteristic of CSM ejected from the progenitor system, is needed. However, this would produce time-varying absorption that is only seen in a few SNe~Ia \citep{Patat07,Simon09,Ferretti16}, whereas the majority of our sample does not show any evolution \citepalias{GG24}. As the absorption of \naid\ in young-HV SNe~Ia is generally higher than for the rest of SNe~Ia, in the ionisation scenario, this means that the sodium quickly ionised (making the absorption decrease) and already recombined (making the absorption increase back) even before maximum, since many of our spectra are pre-maximum. This would indicate extremely nearby material with very high electron density. Nonetheless, if we assume that both channels, young-NV and young-HV, come from the same progenitor system, with presumably the same CSM material, then the total amount of sodium before the ionisation and after the recombination should be preserved, so that both channels should have the same sodium column density after recombination. This signifies that either the CSM is extremely asymmetric and somehow aligns with explosion asymmetry (being therefore different for young-HV and young-NV), or some other process is responsible for the larger abundance of sodium seen in young-HV SNe. In the next section, we present an ISM interpretation capable of explaining these findings.

\subsection{Grain disruption by SN radiation}\label{sec:RAT}

SN radiation can interact with nearby IS dust clouds and alter the dust size distribution responsible for the $R_V$ via a) dust sublimation, but the radius ($<0.2$pc) where this occurs is too small \citep{Waxman00,Hoang19}, b) cloud-cloud collisions \citep{Hoang17}, but the duration for grain shattering ($\sim100$ yr) is too long \citep{Hoang19}, and c) radiative torques that generate centrifugal forces leading to dust 
disruption \citep{Hoang19,Giang20}. The latter two processes are particularly interesting because they can also release sodium and other species trapped in dust grains \citep[dust depletion, see e.g.,][]{Field74,Konstantopoulou22} back into the gas-phase. In such cases, sodium abundance is increased without a corresponding increase in extinction, as is observed for the young-HV sample (Fig.~\ref{fig:3clust_EW-EBV}). Moreover, in a patchy ISM, multiple cloud components can increase the EW of sodium without an increase in extinction \citep{Maxted25}.

In the asymmetric \subCh\ DDet scenario, the events with initial surface detonations facing the observer correspond to our group of young-HV objects. These also have higher extinction and stronger \naid\ EW absorption. If radiative torques from the SN radiation are responsible for disrupting the grains, shifting their sizes (and $R_V$) and making sodium more abundant, then that SN group needs to have an additional early source of radiation capable of imprinting that extra radiation pressure that, by maximum light, has already changed the grain distribution. In principle, the helium burning in the outer shell of the WD could provide the additional radiation at early times, or even shocks from the collision between the CO detonation and the He detonation \citep{Piro25}. In fact, some SNe~Ia do show signs of early emission \citep[e.g.,][]{Marion16,Hosseinzadeh17,Dimitriadis19,Ni23} with variations in their early-time light-curves and colours \citep{Bulla20,Han20}, possibly making up two groups that relate to ejecta velocities \citep{Stritzinger18}. Interestingly, \citet{Ni25} recently provided evidence for three populations based on early colours, whereby one population would come from old environments, and the two others have more similar young environments. 

The radiation pressure of strong radiation fields of bolometric luminosity, $L_{\mathrm{bol}}$, and effective wavelength, $\lambda_{\mathrm{eff}}$, can accelerate nearby clouds \citep{Hoang17} with grains of size, $a$, located at distance, $d$, to terminal velocities, $v$, of:

\vspace*{-4mm}
\begin{equation}
v \simeq 171\left(\frac{L_{\mathrm{bol}}}{10^8L_{\odot}}\right)^{1/2}\left(\frac{d}{100\mathrm{pc}}\right)^{-1/2}\left(\frac{a}{10^{-5}\mathrm{cm}}\right)^{-1/2}\frac{\mathrm{km}}{\mathrm{s}}. 
\end{equation}

\noindent The bolometric luminosity of the early excess He detonation depends on the WD mass and the He layer mass \citep{Polin19}, but it can reach $10^8L_{\odot}$ to imprint $\sim$100 \kms\ to the clouds, as has been measured for the \naid\ blueshifts. 

Additionally, the radiative torque (RAT) model predicts that the irregular grains can be spun up to fast angular speeds, $\omega$, as follows: 

\vspace*{-5mm}
{\small
\begin{equation}
    \omega = A \left(\frac{a}{10^{-5}\mathrm{cm}}\right)^n\left(\frac{\lambda_{\mathrm{eff}}}{5\mathrm{\mu m}}\right)^{-1.7} \left(\frac{L_{\mathrm{bol}}}{10^9L_{\odot}}e^{-\tau}\right)^{1/3}\left(\frac{d}{\mathrm{pc}}\right)^{-2/3}\frac{\mathrm{rad}}{\mathrm{s}}, 
\end{equation}
}

\noindent where $A$ and $n$ depend on the wavelength and grain size (see \citealt{Hoang19}). For a SN, the evolution of the emission needs to be solved to obtain $\omega(t)$. The centrifugal forces created by the spin-up can tear the grains apart if they exceed a critical rotation of $\omega_{\mathrm{disr}}=
\frac{2}{a}\left(\frac{S_{\mathrm{max}}}{\rho}\right)^{1/2}$, where $\rho$ is the grain mass density and $S_{\mathrm{max}}$ is the maximum tensile strength of the grain, which is rather uncertain for dust. \citet{Giang20} show that grains with maximum tensile values between $10^7-10^{10}$ erg cm$^{-3}$ can be disrupted 10 to 40 days from explosion for clouds at $d=1$ pc due to the SN radiation alone. Excess emission from the surface detonation can destroy them earlier and at larger distances. Interestingly, maximum tensile strengths decrease for larger, composite grains, i.e. those are easier to disintegrate. This means that grains that have metal species attached are possibly easier to destroy, and the fraction of excess sodium SNe~Ia should, in principle, be higher than the fraction of low-$R_V$ SNe~Ia. Taking our young-HV sample, we measure a fraction of 67\% of SNe with $R_V$ values lower than 2.6, which is the lower limit of the MW range. This fraction is higher than the objects with excess sodium (51\%), but the statistics are too low to draw firm conclusions. 

The RAT mechanism can release "new" free sodium disrupting metals attached to dust grains to much farther distances from the SN than what ionisation of neutral sodium from UV photons can do. This is because the RAT process depends mainly on the SN radiation strength, i.e. on the bolometric luminosity, and not just on a narrow UV band in which a lot of lines are blocked by heavy elements. Moreover, the observed increase in column density requires posterior recombination after ionisation, which can only happen at high densities, while the RAT can disrupt grains in more diffuse regions farther from the explosions. Finally, since radiative torques increase with grain size, the RAT disruption is particularly efficient for large grains (such as the ones holding metals like sodium), reaching thus larger distances.

It is interesting to note here that, as opposed to geometrical effects, the reach and strength of the RAT depend on the metal species being released (through its tensile strength parameter) just as the ionisation changes for each species (through the ionisation potential). We find that the EW distributions found for other narrow lines also change among the three groups (see section~\ref{sec:other-lines}), notably for \ion{Ca}{ii} H\&K and \ki\ 1 (which have higher and lower ionisation potential than sodium, respectively), but not for \ki\ 2 nor the DIBs. Naturally, one of the two lines of a doublet is stronger (e.g. \caii\ K is stronger than H, as is \ki 1 vs 2), and the ratio can tell us about the optical depth of the medium. If the ratio nears one, the lines are becoming saturated, and the medium is optically thick, which is not the case on average for our sample. The DIBs are complex, unknown molecules that correlate with dust extinction \citep[e.g.,][]{Phillips13}; however, they do not show any evolution, and we do not find differences within our groups. Since all these lines are generally weaker and harder to measure, it is difficult to draw firm conclusions.

One problem of the RAT disruption model, as presented in \citet{Hoang19} and \citet{Giang20}, is that all SNe~Ia would end up sooner or later disrupting grains (even without early emission), and even core-collapse SNe (CC-SNe) would as well. The evidence for nearby material in SNe~Ia comprises only a fraction of them, and not all CC-SNe show evidence of intervening matter. In fact, some CC-SNe have surprisingly less sodium absorption than expected for their star-forming regions \citepalias{G25}. Given the uncertainty in some of the model parameters, particularly the tensile strength, there is room for improvement. 

\begin{figure}
\centering
\includegraphics[width=0.95\columnwidth]{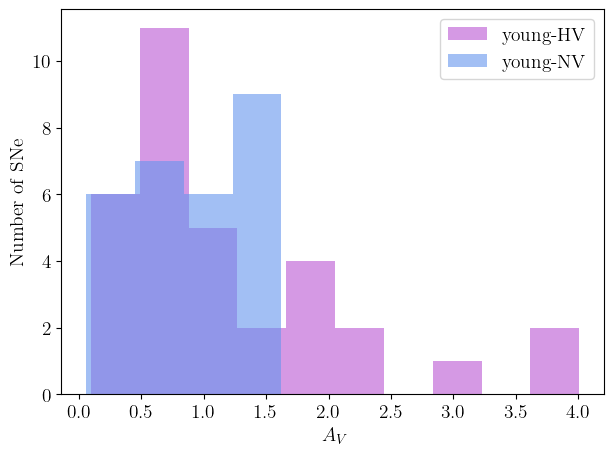}
\vspace*{-4mm}
\caption{$A_V$ distributions for the young-NV and young-HV SNe~Ia.
}
\label{fig:3clust_Av}
\end{figure}
\vspace*{-0.4mm}

An important consideration in the RAT disruption model is that the dust mass around the young population (HV and NV) should, on average, be equal because of its ISM origin. This implies that the additional redder colours seen in the young-HV objects should be explained solely by changes in the extinction curve and by dust scattering, without changing the dust mass. Qualitatively, this could be explained through dust grains becoming smaller (from $R_V$ of 3.1 to 1.5), and preferentially absorbing and scattering wavelengths closer to $B$ and $V$. In fact, \citet[][see their figs.~3 and 4]{Giang20} present this change in the extinction curve for several cloud distances and tensile strengths, assuming dust mass conservation. They predict in all cases that $A_\lambda$ augments at $\lambda<0.4\mu$m while it decreases at $\lambda>0.4\mu$m. At the wavelength of $V$-band, comparing the median value of the young-NV population of $<A_V>=<R_V>\times <EBV>=1.0$ with that of the young-HV $<A_V>=0.79$, we see that it decreases accordingly; however, for the most extincted objects, there are no equivalents with higher extinction in the young-NV population (see Fig.~\ref{fig:3clust_Av}). This poses a problem for the RAT disruption interstellar origin of the material under the assumptions of \citet{Giang20}. The scattering effect, for which bluer light from earlier epochs of the SN is scattered by the dust into the LoS, would only decrease $EBV$ (and increase $R_V$ with constant $A_V$\footnote{Private communication with M. Bulla}) of the young-HV SNe~Ia, unable to explain this discrepancy.

\subsection{Cosmological mass-step}\label{sec:mass-step}

\begin{figure}
\centering
\includegraphics[width=0.96\columnwidth]{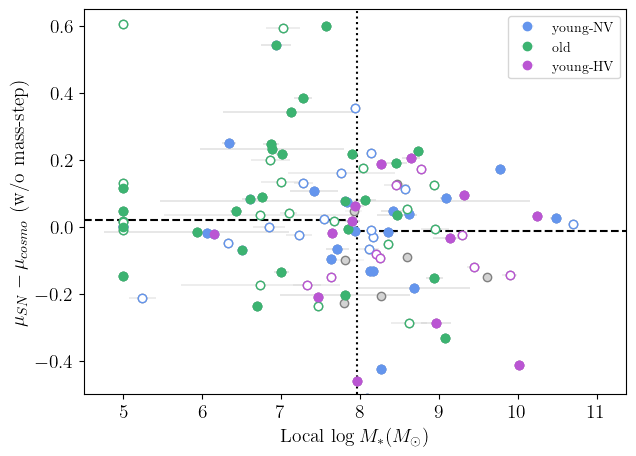}
\includegraphics[width=\columnwidth]{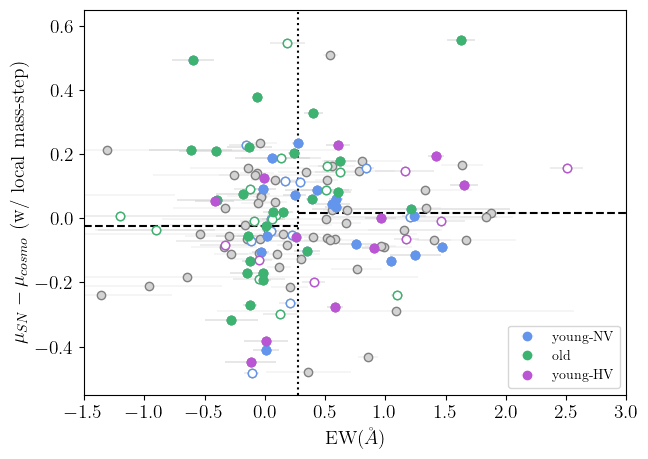}
\vspace*{-7mm}
\caption{\emph{Upper:} HR without mass-step as a function of local stellar mass. \emph{Lower:} HR with local mass-step correction as a function of \naid\ EW. The three GMM clusters are shown in colours for the initial (filled) and imputed samples (open). Median of the entire population is shown as vertical dotted lines. The horizontal dashed lines indicate the median of the HR for low and high EW or mass values, respectively. 
}
\label{fig:HRL-EW}
\end{figure}

The mass-step in SN~Ia cosmology is the third calibration correction, and it depends on the host environment. Its origin is actively debated in the literature with claims for an extrinsic dust origin, an intrinsic SN difference or a combination of both \citep[e.g.,][]{Johansson21,Thorp21,Meldorf23,Popovic24}. In the light of the current study, if the young and old channels are really two independent SN populations with possible different powering mechanisms, then the brightness-stretch correction, $\alpha$,  could be different (see Sect.~\ref{sec:double}), affecting the calibration \citep{Ginolin25}. If we assume that these two populations have the same Phillips relation apt for calibration, then we arrive at the brightness-colour correction, $\beta$, that is composed of an intrinsic colour and a dust extinction factor. Let us suppose that the intrinsic colours and their relation to intrinsic luminosity are equal for all SN~Ia groups. Moreover, let us assume that reddening and absorption due to dust follow the same linear trend as the intrinsic colour-brightness relation. Under these assumptions, we still have the problem of a varying amount of dust: the young population, particularly the HV one, suffers from more dust absorption in $B$ and $V$ bands than the old sample. This is a principle, not an issue, since this second colour calibration would take that into account. However, $\beta\simeq R_V+1$ and we show that $R_V$ changes substantially for the two young sub-samples, invalidating therefore a universal $\beta$ calibration. To exacerbate things, dust scattering and echoes from this nearby material (present especially in young-HV SNe~Ia) can add blue light from earlier emission into the maximum SN light, depending on the geometry and distance of the clouds \citep{Bulla18}, and going in the opposite direction of the reddening-extinction relation.
So, even in the most optimistic scenarios that fulfil many of our assumptions, the simplistic SN~Ia calibration seems unreasonable and dangerous for a large and heterogeneous population. Multiple channels can have important implications for cosmological constraints \citep[e.g.,][]{Wojtak25,Martins25}.

A third calibration may, in fact, be compensating for over-corrections to populations from the other two calibrations.  In upper Fig.~\ref{fig:HRL-EW}, we show the HR (without mass-step correction) as a function of the local stellar mass. The low-mass objects on the left make up mostly the old population, whereas the high-mass objects correspond to the young one. In that sense, the mass-step would be naturally explained as a split of two populations that have intrinsic differences (e.g., in asymmetries) but also significantly different environments, such as stellar age and dust content. These findings agree with studies showing that the mass-step is a result of local differences such as stellar age \citep{Rose19,Rigault20,Sarin26}. In a recent work, \citet{Burgaz25} show that NV have a strong mass-step, while for HV objects it is consistent with zero. If NV are indeed partly split into the old and young components, this is consistent with our results: NV would have two intrinsic populations that the mass-step picks up. If HV objects are part of a single population, on the other hand, then it also makes sense that they have less scatter in the HR \citep{Wang09}. \citet{Toy25} find that the mass-step in the outskirts is reduced with respect to the central parts of the galaxies. As a single channel, the old population is preferentially located in the outskirts, while in the centre we have the two young populations with differing dust properties, so it makes sense that no mass-step is present for a single homogeneous old and outer population. However, a strong mass-step in the inner region means a relation to mass and environment. We confirm in Tab.~\ref{tab:KSdoub} that two HR distributions within the young population have a strong environmental dependence (FF$_{\mathrm{env}}\sim$1e-3) unrelated to dust (no EW difference), perhaps indicating changes in intrinsic luminosity for different progenitor ages, e.g., different WD mass in the \subCh\ scenario (see Sect.~\ref{sec:2pops}). 

In Tab.~\ref{tab:KSsimp} we find that the EW distributions change according to HR. The difference is stronger for HR after the local mass-step correction.
In Fig.~\ref{fig:HRL-EW} we show the HR as a function of EW, and we highlight the median HR differences between low-EW and high-EW objects (below and above <EW>=0.203). This division by the median separates most of the old (with low EW) from the young (with high EW) SNe. It indicates that even though the mass-step correction was included, the young population is fainter after all corrections. By correcting with a higher assumed $\beta$ (i.e. a higher $R_V$) we over-correct the young-HV population that ends up being fainter. Some of the SNe with the highest EW and HR in the figure are indeed from the young-HV sample. However, we do not see in Tab.~\ref{tab:3clust} that the young-HV HR are significantly fainter than the young-NV, as one would expect under this hypothesis. Larger samples will shed more light.

Finally, although a fourth standardisation correction could be included to further diminish HR based on the EW, we strongly advocate not to do this and rather focus on well-understood homogeneous sub-populations of SNe~Ia for cosmology.

\section{Conclusions}
\label{sec:conc}

In this work, we have analysed the EW and VEL of the narrow \naid\ absorption with respect to a variety of properties from the hosts and the SNe, both photometric and spectroscopic. Our main conclusions are:

\begin{itemize}
    \item The largest driver of the strength of sodium in the LoS of SNe~Ia is the environment:  SNe in young, star-forming, central regions present significantly more \naid\ absorption than those in older, more quiescent, outer regions.
    \item The observed colours at maximum and at late times (60 days post maximum), as well as the colour excess of SNe~Ia, are strongly related to \naid\ abundance, confirming that it is a good tracer of extinction. The light-curve width parameters are also significantly different according to the EW, which is expected given their environmental dependence.   
    \item Nebular velocities are strongly related to EW of \naid\ showing that there is a relation between intrinsic SN properties and the gas and dust in the LoS.
    \item We find an optimal environmental division between two SN~Ia samples (based on sSFR) and show that the young sample has important differences in the EW of \naid\ when divided according to colours, colour excess, $R_V$, silicon ejecta velocity, silicon velocity gradient, and nebular velocity. This strengthens the idea that within the same young local environment, either intrinsic SN properties affect the intervening material, or immediate environmental properties (at $r<0.5$ kpc), which correlate with the abundance of intervening material, influence the SN explosion. 
    \item The young-HV channel, besides being composed of HV objects with low $R_V$ and high extinction in the $V$ band, is also consistent with more blueshifted \naid\ lines and excess absorption with respect to what is expected from its extinction.
    \item We provide a possible interpretation of these results in which the young population comes from an asymmetric explosion that interacts with nearby material. In the \subCh\ DDet, for example, early emission on the side of the first detonation could imprint additional radiation that pushes nearby ISM clouds and spins up dust grains through radiation torques, disrupting them, lowering the size distribution ($R_V$) and releasing metals trapped in grains to the gas-phase. Nonetheless, details of this scenario need to be properly addressed, e.g. how and when the disruption occurs and why it affects only a fraction of SNe~Ia, or why there seem to be deviations in the expected $A_V$ distributions for two orientations of the young populations subject to the same ISM dust mass.  
    \item The local mass-step in cosmology can be explained through these two, old and young, populations occurring in different mass regimes, as well as in different offsets from the galaxy centres. Aside from differences between the two populations in intrinsic properties, such as luminosity, colour and the luminosity-stretch and colour relations, one of the populations has clearly a fraction of SNe~Ia with lower $R_V$ (the young-HV) that, when corrected with a single higher $\beta$ parameter, get fainter after standardisation.  
\end{itemize}

This work provides further light into the SN~Ia puzzle, showing that these objects are complex and heterogeneous, that the intrinsic properties are related to nearby material, and that cosmological standardisation should be carefully revisited. 

\begin{acknowledgements}
We thank the anonymous referee for the comments and suggestions that have helped us to improve the paper.
We acknowledge the financial support from the Mar\'ia de Maeztu Thematic Core at ICE-CSIC. 
C.P.G. acknowledges financial support from the Secretary of Universities and Research (Government of Catalonia) and by the Horizon 2020 Research and Innovation Programme of the European Union under the Marie Sk\l{}odowska-Curie and the Beatriu de Pin\'os 2021 BP 00168 programme. 
C.P.G. and L.G. recognise the support from the Spanish Ministerio de Ciencia e Innovaci\'on (MCIN) and the Agencia Estatal de Investigaci\'on (AEI) 10.13039/501100011033 under the PID2023-151307NB-I00 SNNEXT project, from Centro Superior de Investigaciones Cient\'ificas (CSIC) under the PIE project 20215AT016 and the program Unidad de Excelencia Mar\'ia de Maeztu CEX2020-001058-M, and from the Departament de Recerca i Universitats de la Generalitat de Catalunya through the 2021-SGR-01270 grant.
J. D., R. S. and G. M. acknowledge support by FCT for CENTRA through the Project No. UID/99/2025. J. D. also acknowledges support by FCT under the PhD grant 2023.01333.BD, with DOI \url{https://doi.org/10.54499/2023.01333.BD.}

This research has made use of the \textsc{python} packages \textsc{astropy} \citep{astropy:2013,astropy:2018,astropy:2022}, \textsc{numpy} \citep{numpy}, \textsc{matplotlib} \citep{matplotlib}, \textsc{scipy} \citep{scipy}, \textsc{pandas} \citep{pandas,mckinney-pandas}.

\end{acknowledgements}

\bibliographystyle{aa} 
\bibliography{Bibliography}


\begin{appendix}

\section{Details of fits and measurements}\label{ap:details}

In this section, we show in more detail the photometric, spectroscopic and cosmological measurements performed in sections~\ref{sec:phot}, \ref{sec:spec} and \ref{sec:HR}. A full table with all fitted parameters is presented in App.~\ref{ap:longtable}.

\subsection{Photometric fits}\label{ap:details_phot}

\begin{figure}
\centering
\includegraphics[width=0.9\columnwidth]{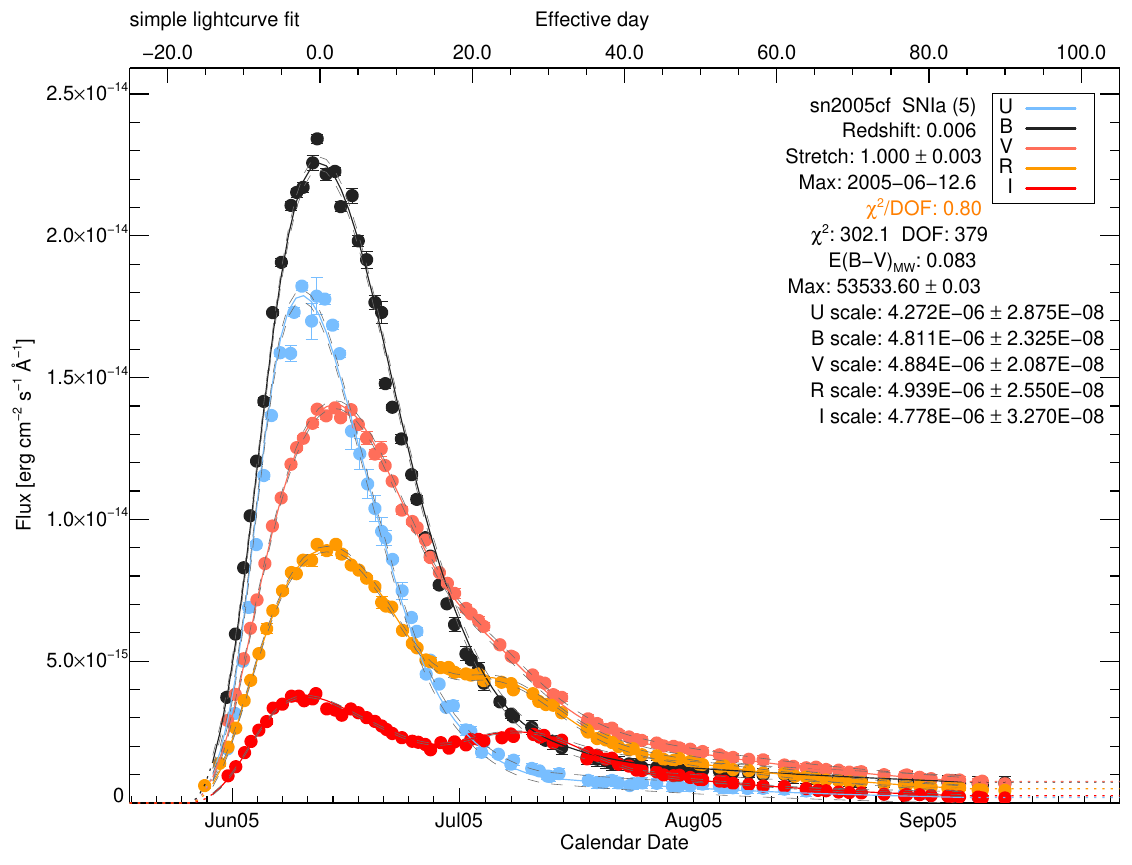}
\includegraphics[width=0.87\columnwidth]{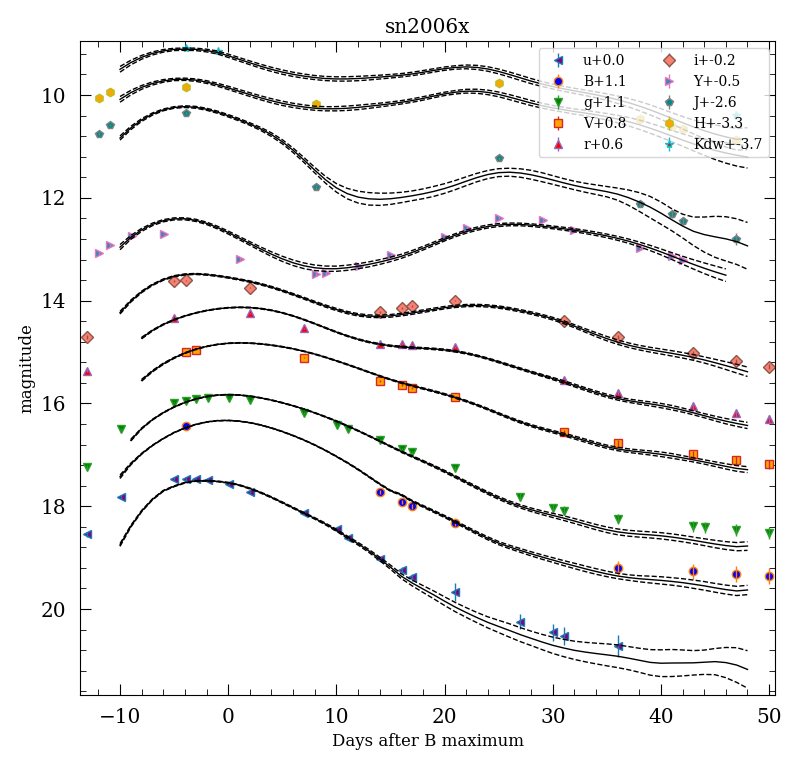}
\vspace*{-3mm}
\caption{Example multi-band light-curve fit with \textsc{SiFTO} (upper) for SN~2005cf in $UBVRI$ flux units and with \textsc{SNooPy} (lower) for SN~2006X in $uBVgriYHK$ magnitudes.
}
\label{fig:LCfits}
\end{figure}

The light-curve fitters \textsc{SiFTO} and \textsc{SNooPy} use spectral template series from \citet{Hsiao07} brought to the observer frame and convolved with the filter transmission to compare the synthetic fluxes with observed multi-band photometry. The time of maximum and the stretch are free parameters. In the case of \textsc{SiFTO}, the stretch parameter $s$ elongates or compresses the time axis of the $B$-band light-curve, spanning values of 0.6 for extremely short-lived 91bg-like objects up to 1.2 for very long-lived, typically 91T-like, SNe~Ia. For \textsc{SNooPy}, the colour-stretch $s_{BV}$ is a measure of the elongation of the colour curve given by the time of its maximum, $s_{BV}=t_{max,BV}/30d$, spanning values from 0.3 to 1.3. The other key difference between the two fitters is that \textsc{SiFTO} is agnostic to the reddening laws, allowing a free scale parameter for each fitted band, whereas \textsc{SNooPy} assumes stretch-dependent intrinsic colours from which a single scale parameter is fitted and the rest are derived from the dust extinction $EBV$ and its wavelength dependence $R_V$. For best performance, \textsc{SNooPy} should use NIR bands to anchor the reddening law. It uses MCMC for posterior sampling, and we use a uniform prior for $s_{BV}$, an exponential decline for $EBV$ with $\tau=0.2$ and a "bin" prior for $R_V$ that changes according to $EBV$ (see \citealt{Burns14}). To ensure proper estimates of the date of maximum, the light-curve width and the colour or extinction parameters, we require at least four data points between $-10$ and $+35$ days from maximum, at least one between $-10$ and $+5$, one between $+5$ and $+20$ and at least two filters with data between $-8$ and $+10$. Example fits are shown in Fig.~\ref{fig:LCfits}.

\begin{figure}
\centering
\includegraphics[width=\columnwidth]{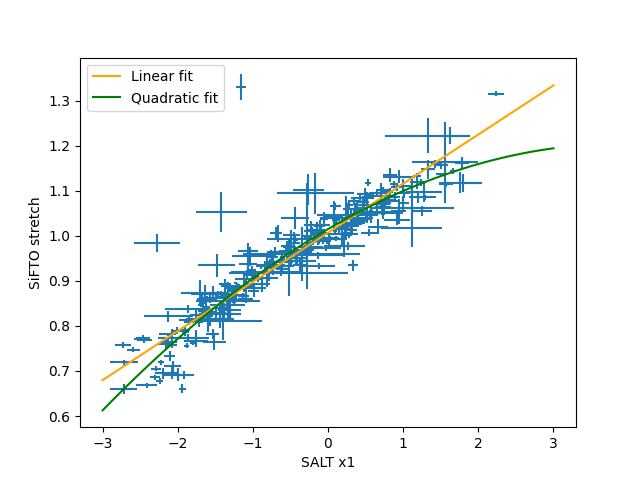}
\vspace*{-7mm}
\caption{Comparison of \textsc{SiFTO} stretch $s$ and \textsc{SALT} $x_1$ for SNe within our sample and Pantheon$+$. Linear and quadratic fits are shown. 
}
\label{fig:SALT}
\end{figure}

In addition to \textsc{SiFTO} and \textsc{SNooPy}, a very popular fitter, especially used in SN~Ia cosmology, is \textsc{SALT} \citep{Guy07,Guy10,Brout22SALT}. We do not perform \textsc{SALT} fits in this study, but we compare the overlapping SNe of the \textsc{Pantheon+} sample \citep{Brout22} in Fig.~\ref{fig:SALT}. In particular, from a linear fit we find that the stretch division for the two populations of $\alpha$ values found in \citet{Ginolin25} of $x_1=-0.48$ corresponds to $s=0.954$, very close to the median value of our sample ($<s> = 0.959$).

\begin{figure}
\centering
\includegraphics[width=\columnwidth]{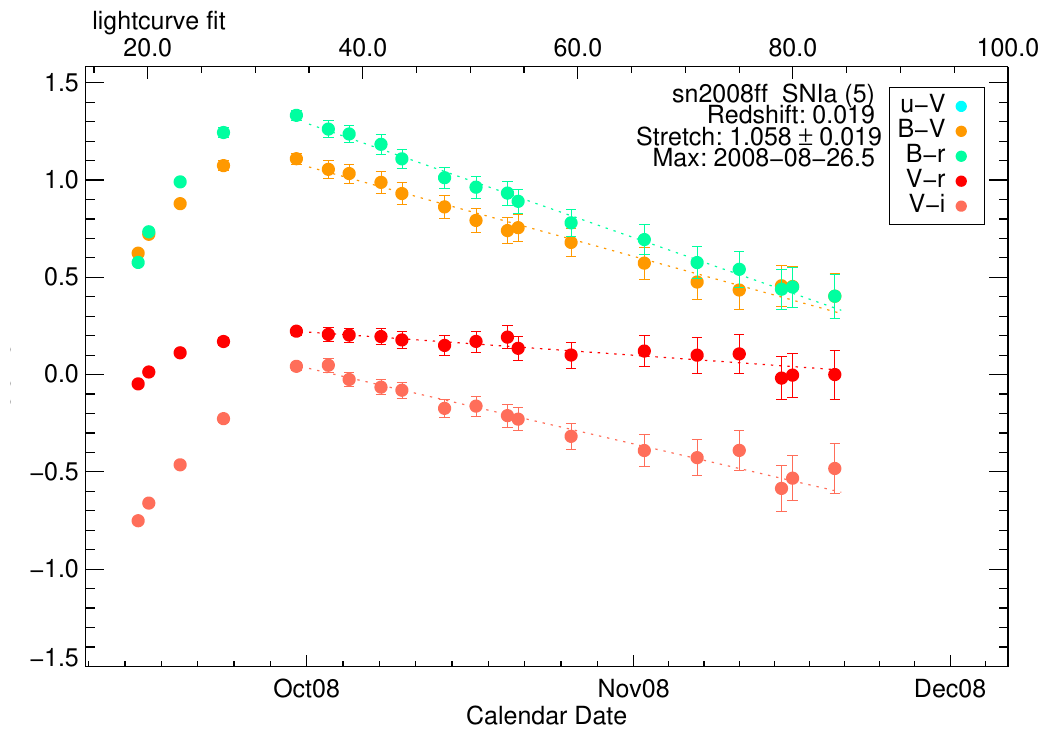}
\vspace*{-7mm}
\caption{Example colour-curve for SN~2008ff with Lira-law linear fits (dashed lines) in $u-B$, $B-V$, $B-r$, $V-r$ and $V-i$. } \label{fig:Lirafits}
\end{figure}

To fit the late-time Lira law, we require at least three simultaneous $B$ and $V$ data points within 35 and 85 days past maximum and a minimum separation of 25 days between the first and last observations. The used photometry is corrected for MW extinction and $K$-corrected to the rest-frame. We perform linear fits during this time interval as shown, for example, in Fig.~\ref{fig:Lirafits}, from which we obtain the Lira slope, $dBV_{60}$ and intercept $BV_{60}$. 

\subsection{Spectroscopic fits}
\label{ap:details_spec}

\begin{figure}
\centering
\includegraphics[width=0.9\columnwidth]{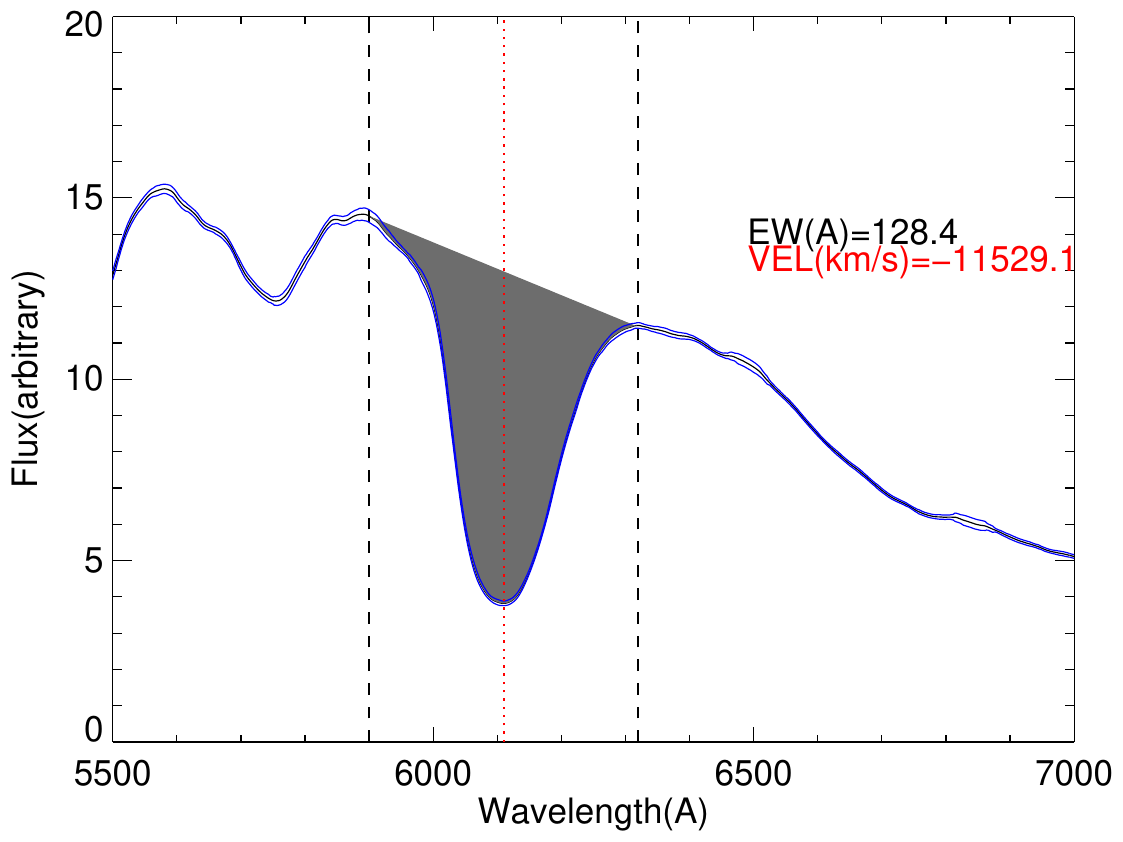}
\includegraphics[width=\columnwidth]{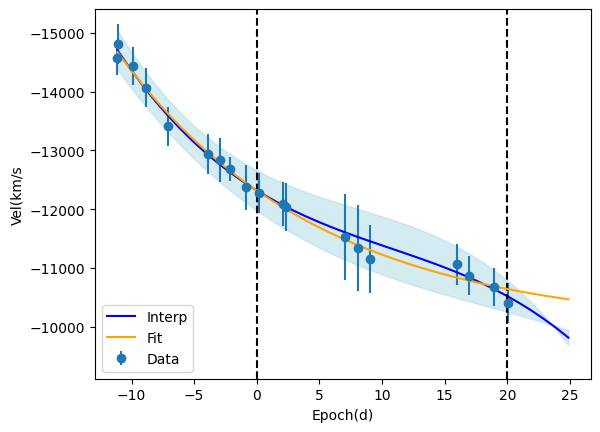}
\vspace*{-7mm}
\caption{\textbf{Upper}: Example smoothed spectrum of SN~2016coj 7 days after maximum around the \ion{Si}{ii}-6355\AA\ line showing the edges of the line (dashed black vertical lines), and the wavelength of the minimum (red dotted line), from which the velocity is calculated. \textbf{Lower}: Evolution of the velocity for the same SN, spline interpolation (blue line) and power-law fit (orange). The two characteristic times (0 and 20 days), from which $v_{grad}$ is calculated, are shown as vertical dashed lines.} \label{fig:velfit}
\end{figure}

To calculate the expansion velocity of \ion{Si}{ii} at 6355\AA, we simply calculate the minimum within a certain region around the line given by the blue- and red-ward edges in \citet{Folatelli13} (see upper Fig.~\ref{fig:velfit}). To avoid spurious extrema from lower S/N, we first smooth the spectrum with a cosine kernel. The measured uncertainty on the velocity comes from the standard deviation of taking the following 10 minima instead. We also add an uncertainty from the difference in calculating the minimum before and after continuum subtraction. We have compared our methodology with the more standard Gaussian-fit approach, finding excellent agreement.

We then use linear or spline interpolation to obtain the velocities at maximum and 20 days after. We require at least one spectrum within $\pm3$ days of maximum and one within $20\pm5$ for each of these two measurements, respectively (see lower Fig.~\ref{fig:velfit}). The uncertainty comes from the interpolation of the data uncertainties. When we only have one velocity data point within the range around maximum and no others to interpolate, we use the average power-law fit found for all SNe~Ia with good coverage (see orange line in the Figure): $v(t)=A\exp(\gamma(t-t_0))+C$, with the fitted values presented in Tab.~\ref{tab:power-law}. We propagate the uncertainties of the parameters for the most extreme high and low velocity gradients shown in the Table.

\begin{table}
\centering
\caption{Velocity evolution power-law average parameters}
\label{tab:power-law}
\renewcommand{\arraystretch}{1.4}
\begin{tabular}{c|ccc}
\hline
\hline
Parameter & Median & Higher & Lower \\
\hline
$A$ & $4.450\times10^{3}$ & $1.146\times10^{4}$ &  $1.730\times10^{6}$\\
$t_0$ & $-1.618 \times10^1$ & $8.397\times10^1$ & $-3.468\times10^3$ \\ 
$\gamma$ & $-0.0666$ & $-0.0155$ & $-0.0523$ \\
$C$ & $8.893\times10^3$ & $-8.114\times10^3$ & $1068\times10^4$ \\
\hline
\end{tabular}
\tablefoot{Median gradient parameters correspond to $v_{grad}\sim60$ \kmsd, higher gradient to $v_{grad}\sim225$ \kmsd\ and lower gradient to $v_{grad}\sim9$ \kmsd.}
\end{table}

\subsection{Cosmological fits}
\label{ap:details_hr}

To obtain the best standardisation parameters with a fixed cosmology, we minimise the following log-likelihood for our $N$ SNe:
$$
\ln(\mathcal{L})=-\frac{1}{2}\sum^{N}_{i=0}\left[ \frac{\mu_{\mathrm{SN},i}-\mu_{\mathrm{mod},i}}{\sigma_i} + \ln(2\pi\sigma_{i}^2 )\right],
$$
\noindent with the uncertainty $\sigma_i$ given only by the diagonal elements of the covariance matrix:

$$
 \sigma_{i}^{2}=\sigma^{2}_{m_{B}}+(\alpha \sigma_{x_1})^{2}+(\beta \sigma_{c})^{2}-2\beta\sigma_{m_{B},c} + 2\alpha\sigma_{m_{B},x_1}
 $$
 $$
-2\alpha\beta\sigma_{x_1,c}-\sigma^{2}_{int}+\sigma^{2}_{lens}+\sigma^{2}_{z}
$$

\noindent where $\sigma_{m_{B}}$, $\sigma_{x_1}$ and $ \sigma_{c}$ are the uncertainties associated to \textsc{SiFTO} parameters of each SN, as well as their respective covariances $\sigma_{m_{B},c}$, $\sigma_{m_{B},x_1}$ and  $\sigma_{x_1,c}$. The parameters $\sigma_{z}$ and $\sigma_{lens}$ correspond to the uncertainty contribution from the redshift uncertainties and from gravitational lensing \citep{Jonsson10}. Finally, $\sigma_{int}$ is a free parameter in the fit that accounts for intrinsic variations in the SN luminosities not captured by the standardisation. 

\begin{figure}
\centering
\includegraphics[width=\columnwidth]{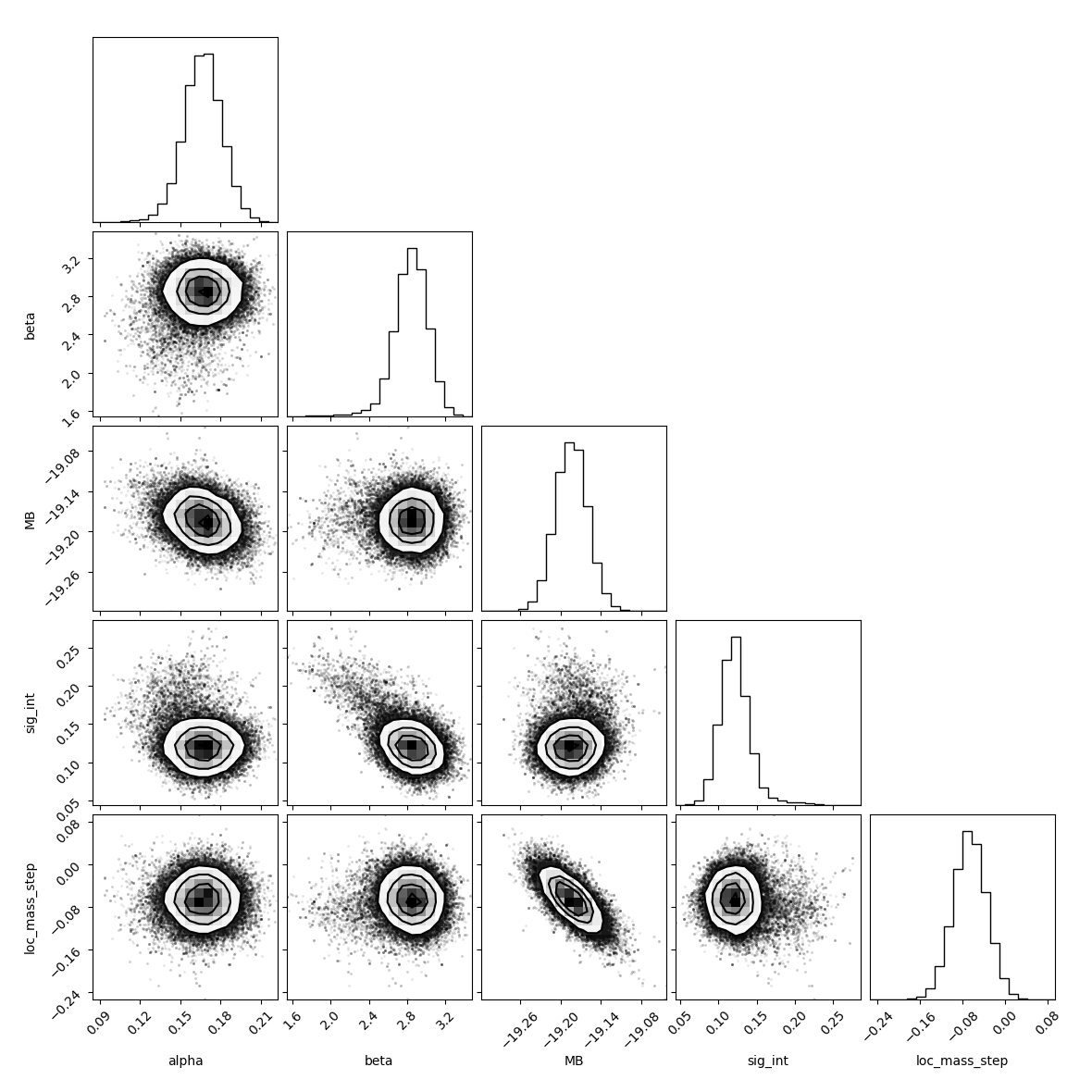}
\vspace*{-7mm}
\caption{Corner plot of the posterior distribution of the fitted parameters: $M_B$, $\alpha$, $\beta$,  $\sigma_{\mathrm{int}}$ and $\Delta_{\mathrm{host}}^L$ for the local stellar mass-step.} \label{fig:corner}
\end{figure}

We use 10 initial optimisations with the Nelder-Mead algorithm \citep{Gao12minimize} implemented in \textsc{scipy}, and then do a posterior sampling with a Markov Chain Monte-Carlo \citep{Goodman10} implemented in \textsc{emcee} \citep{emcee} with 150 walkers and 800 iterations. An example posterior distribution for the local mass case is shown in Fig.~\ref{fig:corner}. The final median and $1\sigma$ uncertainties of the posterior distributions are presented in Tab.~\ref{tab:cosmo}. The recovered mass-step parameters are 2.2 and 2.9$\sigma$ significant for local and global masses, and the root-mean-square (RMS) of the residuals indicates that a model with global mass-step is slightly preferred. It is important to mention that our very nearby low-redshift sample is biased towards targeted large galaxies (as also seen in the large median stellar mass value of $\log M_*^G=10.84$ compared to 10.0 for typical cosmological samples). 

\begin{figure}
\centering
\includegraphics[width=\columnwidth]{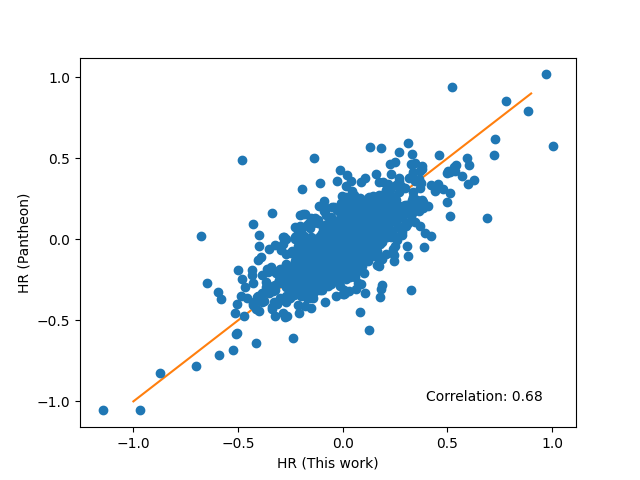}
\vspace*{-7mm}
\caption{Comparison of HR obtained in this work with HR from Pantheon+ \citep{Brout22}, both without mass-step correction.} \label{fig:compHR}
\end{figure}

A more robust cosmological analysis should include the bias corrections \citep[e.g][]{Kessler19} --at least the redshift-dependent Malmquist bias--, as well as non-diagonal covariance elements \citep[e.g.,][]{Conley11}. We stress that the goal of the present study is to evaluate how the HR residuals depend on the narrow line properties of intervening material in the LoS. As such, the exact values of the fitted parameters, as well as the absolute HR values and their median, are irrelevant; we only care about relative differences between subsamples of SNe~Ia. In Fig.~\ref{fig:compHR}, we show that our HR and those of \citet{Brout22} without mass-step correction are comparable despite differences in the fitted parameters.  

\section{Details of populations studies}\label{ap:details}

In this section, we expand on the population studies divided according to environment presented in Sect.~\ref{sec:double}, and the clustering analysis presented in Sect.~\ref{sec:clust}.

\subsection{Environmental subsamples}
\label{ap:FFenv}

\begin{figure}
\centering
\includegraphics[width=\columnwidth]{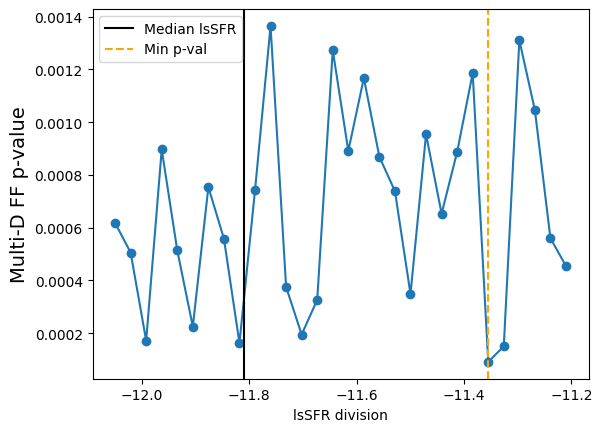}
\vspace*{-7mm}
\caption{Multi-dimensional FF test p-value of local environmental properties (sSFR$^L$, $t_{\mathrm{age}}^L$, $A_V^L$ and $M_*^L$) as a function of local sSFR division. The median of the sample is shown in black and the sSFR for the minimum p-value in orange.} 
\label{fig:multiKS-lsSFR}
\end{figure}

To obtain two sub-samples of SNe~Ia that are significantly different in their environmental properties, we sweep each of the five local properties (sSFR$^L$, SFR$^L$, $t_{\mathrm{age}}^L$, $A_V^L$ and M$_*^L$) within 35 and 65 percentiles of their distributions and calculate for each division the multi-dimensional FF test on the five parameters. An example for the sSFR is shown in Fig.~\ref{fig:multiKS-lsSFR}. The minimum p-value is shown with an orange line and occurs at $\log$ sSFR $= -11.3$. Repeating the search with all other environmental parameters as divisions to the sample results in larger p-values (see Tab.~\ref{tab:FFenv}). As seen in the Figure, there is no real global minimum, and the exact division between the two samples is somewhat arbitrary.

\begin{table}
\centering
\caption{Minimum FF test p-value of local environmental properties}
\label{tab:FFenv}
\renewcommand{\arraystretch}{1.4}
\begin{tabular}{c|ccccc}
\hline
\hline
Property & sSFR$^L$ & SFR$^L$ & $t_{\mathrm{age}}^L$ & $A_V^L$ & $M_*^L$ \\
\hline
Min p-value & 9e-5 & 2e-4 & 3e-4 & 3e-4 &  8e-4\\
\hline
Division & $-11.3$ & $-3.60$ & 1.58 & 2.51 & 8.12\\
\hline
\end{tabular}
\tablefoot{The division values are logarithmic for sSFR, SFR and $M_*$.}
\end{table}

\subsection{SFR-corrected distributions}
\label{ap:SFRcorr}

\begin{figure}
\centering
\includegraphics[width=\columnwidth]{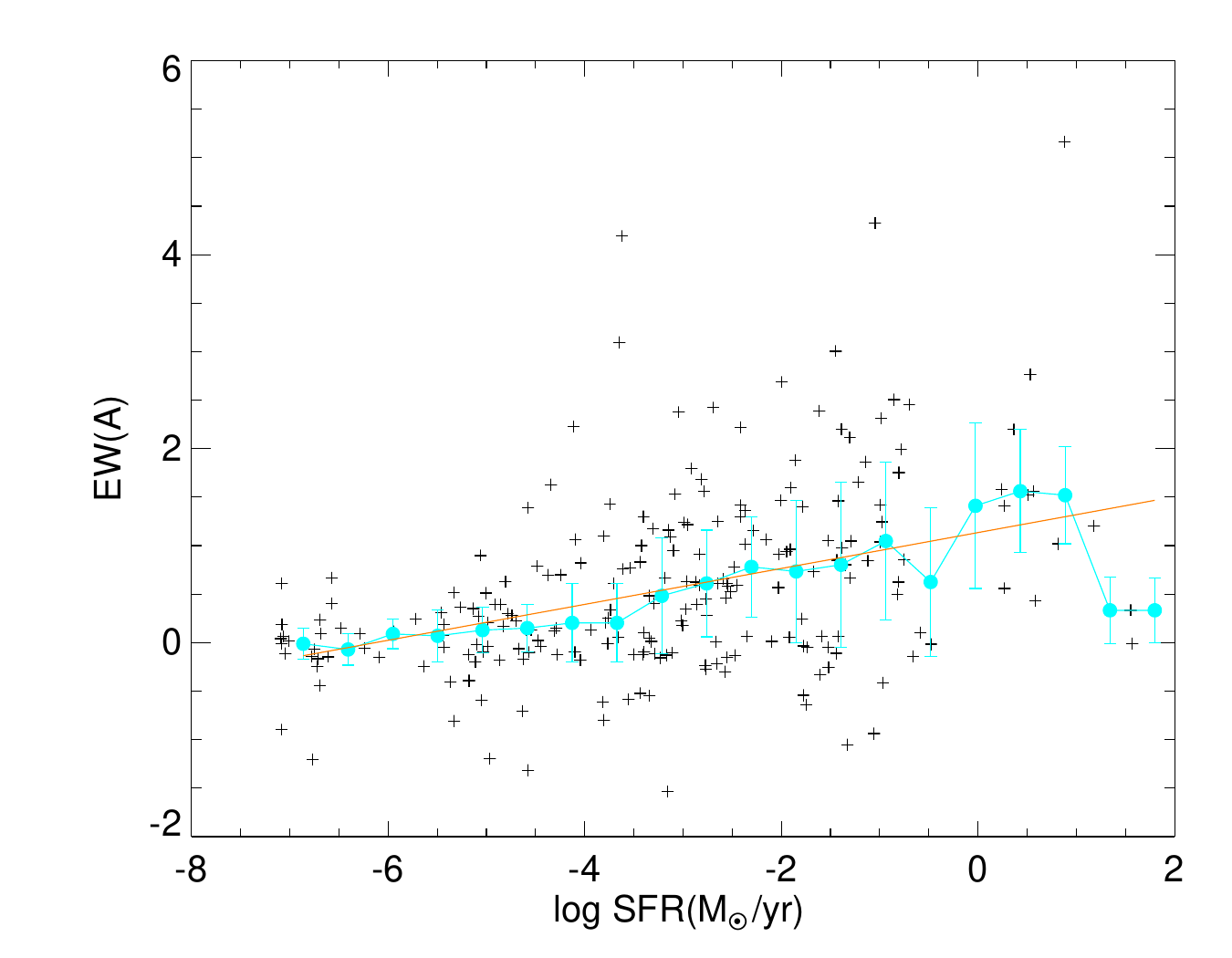}
\vspace*{-7mm}
\caption{\naid\ EW vs $\log$ SFR for SNe (crosses). Median and absolute deviations are shown as cyan circles with error bars, and the orange line represents the best linear fit.}  
\label{fig:sfr-corr}
\end{figure}

To further demonstrate that the EW of \naid\ depends on intrinsic SN properties beyond the local environments, we correct here the measured EW for the SFR dependence in a similar way to \citetalias{G25}. In Fig.~\ref{fig:sfr-corr} we show the dependence of the EW with SFR that is well fitted by an exponential of the form: EW(SFR)$= a\times\log\mathrm{SFR}+b$, with $a=1.13\pm0.16$ and $b=0.19\pm0.03$. By dividing this dependence of the SFR, we obtain a corrected EW$_{\mathrm{corr}}=$EW/EW(SFR), for which we repeat the analysis presented in Sec.~\ref{sec:double} and Tab.~\ref{tab:KSdoub}. We confirm the strong differences for all relevant properties such as:  $EBV$, $R_V$, $v_{max}$ and $v_{neb}$, and reproduce those in Tab.~\ref{tab:KSdoub_corr} for the young population. The median EW differences between the two SN distributions are even larger, and their KS p-values are even stronger (except for $v_{neb}$), confirming that differences in abundance relate to SN properties beyond environments. 

\begin{table}
\small
\centering
\caption{KS tests of two EW$_{\mathrm{corr}}$ distributions (corrected by local SFR) divided according to a single property for the young population}
\label{tab:KSdoub_corr}
\renewcommand{\arraystretch}{1.3}
\begin{tabular}
{C{1.00cm}|C{1.31cm}C{1.52cm}C{1.04cm}C{0.9cm}C{1.05cm}}
\hline
\hline
\textbf{Sample} & \multicolumn{4} {c} {\textbf{YOUNG}} \\
\hline
\textbf{Prop.} & $<\mathbf{EW}_{\mathrm{hi}}>$ & 
$<\mathbf{EW}_{\mathrm{lo}}>$ & \textbf{KS} &$\mathbf{P_{MC}}$ \\ 
\hline
$EBV$  & 1.49$\pm$0.60 & 0.47$\pm$0.51 & 3.1e-4 & 81 \\
$R_V$  & 0.35$\pm$0.54 & 1.68$\pm$0.69 &  7.7e-5  & 99 \\
$v_{max}$ & 0.35$\pm$0.69 & 1.89$\pm$0.48 & 2.5e-3 & 90 \\
$v_{neb}$  &  1.49$\pm$1.08 & $-0.37\pm$0.08 & 0.08 &  6\\
\hline
\end{tabular}
\tablefoot{Similar to Tab.~\ref{tab:KSdoub} with SFR-corrected EW for a sample of properties.}
\end{table}

\subsection{Clustering subsamples}
\label{ap:clust}

The Gaussian Mixture Model \citep[GMM,][]{Mclachlan00} assumes that the data can be probabilistically represented through a set of normal distributions corresponding to each cluster. We use the \textsc{scikit-learn} implementation \citep{scikit-learn}, which starts with a k-means clustering \citep{Steinhaus56} to initialise the standardised parameters\footnote{Parameters are subtracted by the mean and divided by the standard deviation to give each input feature the same weight.} and then iteratively updates them by fitting the GMM to the data with the Expectation-Maximisation (EM) procedure \citep{Dempster77}. As input features, we take two local environmental variables, SFR$^L$ and $t_{\mathrm{age}}^L$, the stretch $s$, the ejecta velocity $v_{max}$ and the reddening law $R_V$. 

After finding the clusters and the membership probabilities for each SN, we take into account the uncertainties in the input parameters by doing a Monte Carlo that randomly shifts the input variables according to their covariance matrix. A new set of membership probabilities is calculated in each iteration, and the median probability of the cluster for each SN is calculated. The final membership is given by the highest probability. With this procedure, only 5 out of 106 SNe~Ia change their initial membership. 

\begin{figure}
\centering
\includegraphics[width=\columnwidth]{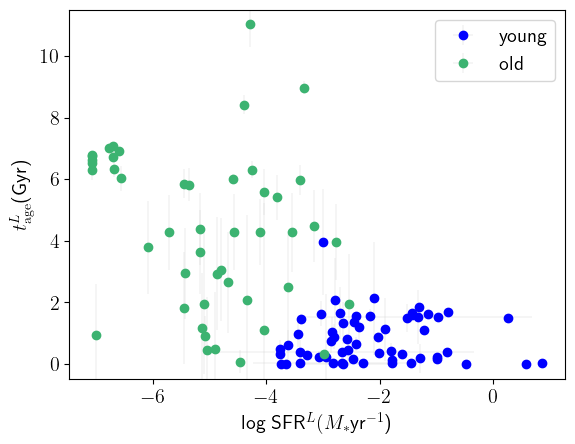}
\vspace*{-7mm}
\caption{Local age vs local sSFR for two GMM clusters: "old" and "young".}  
\label{fig:2clust}
\end{figure}

We perform the analysis for two, three and four clusters\footnote{We also employed another clustering technique for which the number of clusters is not enforced \citep[HDBSCAN,][]{HDBSCAN}, but the number of noise elements encountered by the algorithm was too large.}. For two clusters, the division is strongly related to the local environment, as can be seen in Fig.~\ref{fig:2clust}. Three clusters mostly subdivide the young population into two, as shown in Sect.~\ref{sec:clust}. A fourth cluster only selects a few outliers. The lowest AIC is found for three groups as shown in Tab.~\ref{tab:AIC}.

\begin{table}
\centering
\caption{AIC for GMM clusters}
\label{tab:AIC}
\renewcommand{\arraystretch}{1.4}
\begin{tabular}{c|ccc}
\hline
\hline
GMM & Two & Three & Four \\
\hline
AIC & 1402 & 1391 & 1397 \\
\hline
\end{tabular}
\end{table}

\begin{figure*}
\centering
\includegraphics[width=0.85\textwidth]{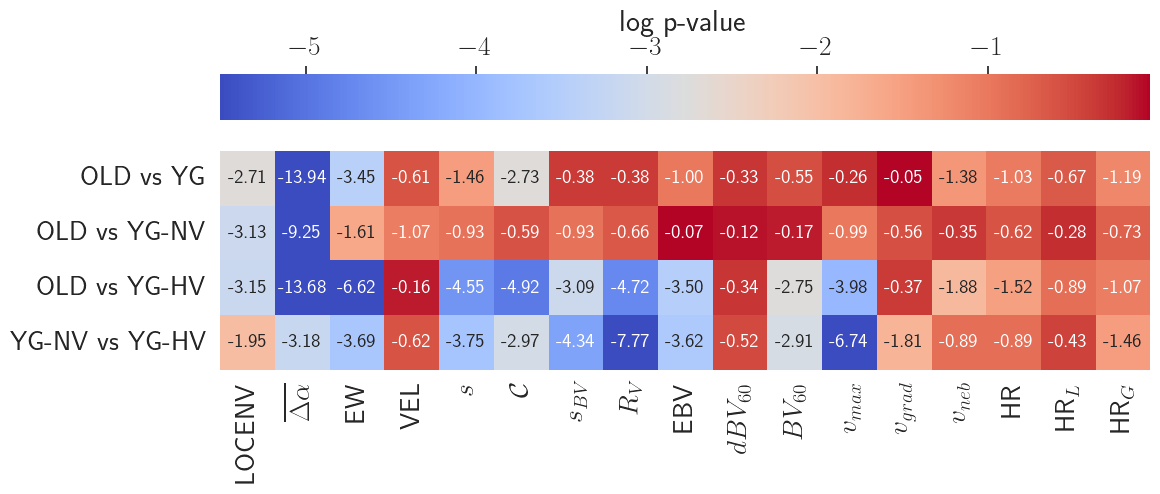}
\caption{Similar to Fig.~\ref{fig:3clustKSmatrix} for the total imputed sample.
}
\label{fig:3clustKSIMPmatrix}
\end{figure*}

The total sample of SNe~Ia used in the clustering consists of 106 objects for which all 5 input features are available. To increase the number of objects in the clustering, we perform a multi-dimensional imputation that replaces missing data with approximate substitute values \citep{Imputation,MultiD-Imputation}. The algorithm is an Iterative Imputation implemented in \textsc{scikit-learn} that does a regression from four features to obtain the fifth feature at each step and for all input features. We only impute SNe for which one of the five features is lacking. More importantly, the imputation is used only to find membership labels for an increased population of 192 SNe, but all the properties shown in Fig.~\ref{fig:3cluster}, as well as the statistics presented in Tab.~\ref{tab:3clustIMP} and Fig.~\ref{fig:3clustKSIMPmatrix} were obtained with real non-imputed values. 

In general, we recover the results of the initial sample with the increased imputed population. Interestingly, the fraction of young-NV SNe~Ia becomes larger (from 47\% to 60\% of the young population), perhaps indicating that these are more frequent. However, the local age difference between young-NV and young-HV SNe~Ia becomes larger and is accompanied also by a decrease in the p-values of the light-curve parameters ($s$ and $s_{BV}$). This could indicate some leakage of older, short-lived SNe~Ia into the young-NV population. Some of the characteristics for the young-HV sample are further strengthened: the $v_{max}$, $v_{grad}$, $R_V$, and all colour properties ($\mathcal{C}$, $EBV$ and $BV_{60}$) have even lower p-values that further differentiate them from the other clusters. 

The results of the EW distributions for other narrow lines are also recovered for the imputed sample (see Tab.~\ref{tab:3clust-lines-IMP}). In fact, in some cases the differences between the young-HV sample and the rest are further strengthened.

\begin{table}
\tiny
\centering
\caption{ Median properties for three SN clusters: "old", "young-NV", and "young-HV" for the imputed sample.}
\label{tab:3clustIMP}
\renewcommand{\arraystretch}{1.3}
\begin{tabular}
{C{1.5cm}|C{1.92cm}|C{1.92cm}|C{1.92cm}}
\hline
\hline
Property & \multicolumn{3}{|c}{Cluster} \\
\hline
 & \textbf{OLD} &\textbf{YOUNG-NV}  & \textbf{YOUNG-HV} \\
\hline
Nr & 69 & 74 & 49 \\
EW & $-0.01\pm0.25$  & $0.27\pm0.38$ & $1.36\pm0.75$ \\
VEL & $-22\pm177$& $127\pm286$ & $-67\pm282$ \\
\hline
$\overline{\Delta\alpha}$ & 0.37$\pm$0.11 & 0.16$\pm$0.07 & 0.11$\pm$0.04 \\
\textbf{SFR}\boldsymbol{$^\mathrm{L}$} & $-5.03\pm$1.06 &  $-2.65\pm$1.14 & $-2.29\pm0.90$ \\
\boldsymbol{$t_{\mathrm{age}}^L$} & 5.72$\pm$1.98 & 0.64$\pm$0.63 & $0.74\pm0.71$ \\
\hline
\boldsymbol{$s$} & 0.91$\pm$0.09 & 0.93$\pm$0.12 & 0.98$\pm$0.05 \\
$\mathcal{C}$ & $0.05\pm0.07$ & $0.09\pm0.10$ & $0.24\pm0.18$ \\
$s_{BV}$  & $0.90\pm0.09$ & $0.88\pm0.16$ & $0.96\pm0.07$\\
$EBV$  & $0.20\pm0.10$ & $0.21\pm0.10$ & $0.37\pm0.18$ \\
\boldsymbol{$R_V$}  & 3.89$\pm$1.55 & 4.29$\pm$1.03 & 2.08$\pm$0.62\\
$dBV_{60}$ & $-0.012\pm0.002$ & $-0.011\pm0.003$ & $-0.012\pm0.003$ \\
$BV_{60}$ & $0.84\pm0.12$ &$0.83\pm0.14$ & $1.06\pm0.15$\\ 
\hline
\boldsymbol{$v_{max}$} & $-10761\pm573$ & $-10642\pm388$ & $-12351\pm1064$\\
$v_{grad}$ & $69.6\pm29.2$ & $69.3\pm21.7$ & $96.0\pm48.6$ \\
$v_{neb}$ & $-119\pm639$& $621\pm776$ & $1905\pm784$ \\
\hline
HR & 0.05$\pm$0.14 & $0.00\pm0.08$ & $-0.03\pm0.12$\\
HR$_{L}$ & 0.04$\pm$0.13 & $0.01\pm0.08$ & $-0.06\pm0.14$ \\ 
HR$_{G}$ & 0.06$\pm$0.11 & $0.02\pm0.06$ & $-0.07\pm0.12$ \\
\hline
\end{tabular}
\tablefoot{Property, median and MAD for three clusters: old, young-NV and young-HV for imputed of 192 SNe. The five input properties for the clustering are highlighted in bold.}
\end{table}

\begin{table}
\tiny
\centering
\caption{Median EW line properties for three SN clusters for the imputed sample}
\vspace*{-2mm}
\label{tab:3clust-lines-IMP}
\renewcommand{\arraystretch}{1.3}
\begin{tabular}
{C{1.5cm}|C{1.92cm}|C{1.92cm}|C{1.92cm}}
\hline
\hline
\textbf{EW} & \textbf{OLD} & \textbf{YOUNG-NV}  & \textbf{YOUNG-HV} \\
\hline
\naid\ &$-0.01\pm0.25$ & $0.27\pm0.38$ & $1.36\pm0.75$ \\
\caii\ H &$0.05\pm0.23$ & $0.27\pm0.24$ & $0.44\pm0.34$ \\
\caii\ K &$0.05\pm0.23$ & $0.14\pm0.27$ & $0.42\pm0.44$ \\
\ki\ 1 &$-0.07\pm0.16$ & $-0.13\pm0.19$ & $0.22\pm0.24$ \\
\ki\ 2 &$-0.07\pm0.15$ & $-0.03\pm0.24$ & $0.03\pm0.24$ \\
DIB 5780 &$0.04\pm0.19$ & $0.04\pm0.12$ & $0.11\pm0.14$ \\
DIB 4428 &$0.26\pm0.31$ & $0.06\pm0.19$ & $0.03\pm0.35$ \\
DIB 6283 &$-0.07\pm0.13$ & $-0.05\pm0.19$ & $0.00\pm0.18$ \\
\hline
\end{tabular}
\end{table}

\section{List of properties}
\label{ap:longtable}

We provide in electronic format the full set of properties for all objects used in this study. The table \ref{tab:long} shows the columns of that table. 

\begin{table}
\tiny
\centering
\caption{Column description of electronic table of SN~Ia and host properties.}
\label{tab:long}
\renewcommand{\arraystretch}{1.3}
\begin{tabular}
{C{1.0cm}C{1.92cm}||C{1.0cm}C{1.92cm}}
\hline
\hline
Column & Property & Column & Property \\
\hline
1 & SN name & 23 & $\sigma(\mathcal{C})$ \\
2 & Redshift & 24 & $s_{BV}$ \\
3 & EW & 25 & $\sigma(s_{BV})$ \\
4 & $\sigma(\mathrm{EW})$ & 26 & $EBV$ \\
5 & VEL & 27 & $\sigma_{EBV}$ \\
6 & $\sigma(\mathrm{VEL})$ & 28 & $R_V$ \\
7 & $\overline{\Delta\alpha}$ & 29 & $\sigma(R_V)$ \\
8 & $\log$ sSFR$^L$ & 30 & $dBV_{60}$ \\
9 & $\sigma(\mathrm{\log sSFR}^L)$ & 31 & $\sigma(dBV_{60})$ \\
10 & $\log$ SFR$^L$ & 32 & $BV_{60}$ \\
11 & $\sigma(\mathrm{\log SFR}^L)$ & 33 & $\sigma(BV_{60})$ \\
12 & $t_{\mathrm{age}}^L$ & 34 & $v_{max}$ \\
13 & $\sigma(t_{\mathrm{age}}^L)$ & 35 & $\sigma(v_{max})$ \\
14 & $A_V^L$ & 36 & $v_{grad}$ \\
15 & $\sigma(A_V^L)$ & 37 & $\sigma(v_{grad})$ \\
16 & $\log M_*^L$ & 38 & $v_{neb}$ \\
17 & $\sigma(\log M_*^L)$ & 39 & $\sigma(v_{neb})$ \\
18 & $\log M_*^G$ & 40 & HR \\
19 & $\sigma(\log M_*^G)$ & 41 & HR$_L$ \\
20 & $s$ & 42 & HR$_{G}$ \\
21 & $\sigma(s)$ & 43 & Cluster member$^{\dagger}$\\
22 &  $\mathcal{C}$ & 44 & LC source$^{\ddag}$ \\
\hline
\end{tabular}
\tablefoot{The units of all properties are shown in Tab.~\ref{tab:props}.\\
$\dagger$ Cluster membership of each SN when available; the imputed members are indicated with; $\ast$ (see Sect.~\ref{sec:clust}).\\
$\ddag$ References of the SN photometry used for light-curve fits (see Additional References).\\}
\end{table}

\end{appendix}

\nocitesupp{Brown14}
\nocitesupp{Shappee16}
\nocitesupp{Stahl19}
\nocitesupp{Chen22}
\nocitesupp{Foley18}
\nocitesupp{Ferretti16}
\nocitesupp{Walker15}
\nocitesupp{Scalzo14}
\nocitesupp{Kilpatrick16}
\nocitesupp{Cadonau90}
\nocitesupp{Buta83}
\nocitesupp{Younger85}
\nocitesupp{Benetti91}
\nocitesupp{Younger85}
\nocitesupp{Tsvetkov90}
\nocitesupp{Kimeridze91}
\nocitesupp{Tsvetkov90}
\nocitesupp{Kimeridze91}
\nocitesupp{Hamuy96}
\nocitesupp{Ford93}
\nocitesupp{Lira98}
\nocitesupp{Meikle00}
\nocitesupp{Krisciunas04}
\nocitesupp{Altavilla04}
\nocitesupp{Riess05}
\nocitesupp{Riess99}
\nocitesupp{Salvo01}
\nocitesupp{Jha06}
\nocitesupp{Turatto98}
\nocitesupp{Modjaz01}
\nocitesupp{Ganeshalingam10}
\nocitesupp{Krisciunas00}
\nocitesupp{Phillips06}
\nocitesupp{Krisciunas06}
\nocitesupp{Krisciunas00}
\nocitesupp{Bufano05}
\nocitesupp{Krisciunas01}
\nocitesupp{Stritzinger02}
\nocitesupp{Li01}
\nocitesupp{Candia03}
\nocitesupp{Valentini03}
\nocitesupp{Lair06}
\nocitesupp{Tsvetkov06}
\nocitesupp{Hicken09}
\nocitesupp{Krisciunas11}
\nocitesupp{Leonard05}
\nocitesupp{Benetti04}
\nocitesupp{Krisciunas04}
\nocitesupp{Phillips07}
\nocitesupp{Pignata08}
\nocitesupp{Pignata04}
\nocitesupp{Ganeshalingam12}
\nocitesupp{Elias-Rosa06}
\nocitesupp{Anupama05}
\nocitesupp{Leonard05}
\nocitesupp{Stanishev07}
\nocitesupp{Krisciunas09}
\nocitesupp{Leloudas09}
\nocitesupp{Chakradhari18}
\nocitesupp{Krisciunas17}
\nocitesupp{Pastorello07a}
\nocitesupp{Taubenberger08}
\nocitesupp{Friedman15}
\nocitesupp{Pastorello07b}
\nocitesupp{Sahu08}
\nocitesupp{Wood-Vasey08}
\nocitesupp{Brown09}
\nocitesupp{Drozdov15}
\nocitesupp{Brown14}
\nocitesupp{Hicken12}
\nocitesupp{Zhang10}
\nocitesupp{Milne10}
\nocitesupp{Contreras10}
\nocitesupp{Brown12}
\nocitesupp{Silverman11}
\nocitesupp{Khan11}
\nocitesupp{Gutierrez16}
\nocitesupp{Szalai15}
\nocitesupp{Foley18}
\nocitesupp{Han20}
\nocitesupp{Yang23}
\nocitesupp{Xi22}
\nocitesupp{Tucker21}
\nocitesupp{Singh22}
\nocitesupp{DerKacy23}
\nocitesupp{Jacobson-Galan22}
\nocitesupp{Zhang22}
\nocitesupp{Srivastav23}
\nocitesupp{Scalzo12}
\nocitesupp{Weyant18}
\nocitesupp{Chakradhari14}
\nocitesupp{Yamanaka16}
\nocitesupp{Li22}
\nocitesupp{Foley13}
\nocitesupp{Stritzinger15}
\nocitesupp{Yamanaka15}
\nocitesupp{Maguire13}
\nocitesupp{Ashall21}
\nocitesupp{Ferretti16}
\nocitesupp{Srivastav17}
\nocitesupp{Wyatt21}
\nocitesupp{Cartier17}
\nocitesupp{Nucita17}
\nocitesupp{Magee16}

\bibliographystylesupp{aa}
\bibliographysupp{references}


\label{lastpage}

\end{document}